\documentclass[pre,twocolumn,superscriptaddress,showpacs]{revtex4-1}
\usepackage{graphicx,epsfig,subfigure}
\usepackage{amsbsy}
\usepackage{amssymb}
\usepackage{amsmath}
\usepackage{bm}
\usepackage{lmodern}
\usepackage[T1]{fontenc}
\usepackage[utf8]{inputenc}
\usepackage{hyperref}
\usepackage[bottom]{footmisc}
\usepackage{xcolor}
\usepackage{ulem}
\newcommand{\eqnref}[1]{Eq.~(\ref{#1})}
\newcommand{\figref}[2]{Fig.~\ref{#2}#1}
\newcommand{\beq}{\begin{equation}}
\newcommand{\eeq}{\end{equation}}
\newcommand{\beqa}{\begin{eqnarray}}
\newcommand{\eeqa}{\end{eqnarray}}
\newcommand{\Tr}{\text{Tr}}

\newcommand{\half}{\frac{1}{2}}
\newcommand{\ab}[1]{\left | #1 \right |}
\newcommand{\rk}[1]{\left( #1 \right)}
\newcommand{\sk}[1]{\left[ #1 \right]}
\newcommand{\bk}[1]{\left\lbrace #1 \right\rbrace}
\newcommand{\bra}[1]{\left \langle #1 \right |}
\newcommand{\ket}[1]{\left | #1 \right \rangle}
\newcommand{\av}[1]{\left\langle #1 \right\rangle}

\newcommand{\ii}{\mathrm{i}}
\newcommand{\ee}{\mathrm{e}}
\newcommand{\dd}{\mathrm{d}}

\date{\today}

\begin{document}

\title{Quantum-classical correspondence of work distributions for initial states with quantum coherence}

\author{Rui Pan}
\affiliation{School of Physics, Peking University, Beijing 100871, People's Republic of China}

\author{Zhaoyu Fei}
\affiliation{School of Physics, Peking University, Beijing 100871, People's Republic of China}

\author{Tian Qiu}
\affiliation{School of Physics, Peking University, Beijing 100871, People's Republic of China}

\author{Jing-Ning Zhang}
\affiliation{Center for Quantum Information, Institute for Interdisciplinary Information Sciences, Tsinghua University, Beijing 100084, People's Republic of China}

\author{H. T. Quan}
\email{Corresponding author: htquan@pku.edu.cn}
\affiliation{School of Physics, Peking University, Beijing 100871, People's Republic of China}
\affiliation{Collaborative Innovation Center of Quantum Matter, Beijing 100871, People's Republic of China}

\begin{abstract}
{The standard definition of quantum fluctuating work is based on the two-projective energy measurement, which however does not apply to systems with initial quantum coherence because the first projective energy measurement destroys the initial coherence, and affects the subsequent evolution of the system. To address this issue, several alternative definitions, such as those based on the full counting statistics and the Margenau-Hill distribution, have been proposed recently. These definitions seem ad hoc because justifications for them are still lacking. In the current study, by utilizing the quantum Feynman-Kac formula and the phase space formulation of quantum mechanics, we prove that the leading order of work distributions is equal to the classical work distribution. Thus we prove the validity of the quantum-classical correspondence of work distributions for initial states with quantum coherence, and provide some justification for those definitions of work. We use an exactly solvable model of the linearly dragged harmonic oscillator to demonstrate our main results.
}
\end{abstract}

\maketitle

\section{Introduction}

Traditionally, thermodynamics describes the energy conversions of macroscopic systems, in which thermodynamic variables, such as work, heat, and entropy production are quantities of ensemble average. Fluctuations of these quantities are vanishingly small and are usually ignored. However, at the microscopic scale, fluctuations are too prominent to be ignored. In the past two decades, stochastic thermodynamics emerges as a new field \cite{jarzynski2011equalities, seifert2012stochastic, sekimoto1998langevin, sekimoto2010stochastic, li2010measurement, martinez2016brownian, bang2018experimental}, in which the classical fluctuating work is defined along individual trajectories in the phase space \cite{jarzynski1997nonequilibrium, sekimoto1998langevin, sekimoto2010stochastic}, leading to the striking results of the fluctuation theorems \cite{jarzynski1997nonequilibrium, crooks1999entropy, seifert2005entropy, hummer2001free, kawai2007dissipation, esposito2010detailed, gong2015jarzynski, liphardt2002equilibrium, collin2005verification, douarche2006work, junier2009recovery, toyabe2010experimental, pekola2015review, hoang2018experimental}. Nevertheless, in the \emph{quantum} regime, there is some ambiguity in the definition of the stochastic trajectory and the corresponding quantum work functional \cite{talkner2016aspects, funo2018path, funo2018heat} due to the existence of the uncertainty principle.

In quantum thermodynamics \cite{kurchan2000quantum, tasaki2000jarzynski, talkner2007fluctuation, talkner2016aspects, funo2018path, funo2018heat, PhysRevE.93.062134, PhysRevE.93.062106, PhysRevE.93.012127, PhysRevE.97.012128, PhysRevE.94.062133, PhysRevE.95.012149, PhysRevE.97.062117, PhysRevE.94.062122, PhysRevE.96.042119, PhysRevA.94.012107, PhysRevA.94.042305, PhysRevX.8.011033, silveri2017quantum, frenzel2016quasi, guarnieri2018quantum, kwon2018fluctuation, peng2017perturbative, strasberg2018operational, goold2016role}, 
a standard definition of quantum fluctuating work is given by the energy difference between the initial and the final outcomes of the two-projective energy measurement (TPM) \cite{kurchan2000quantum, tasaki2000jarzynski, talkner2007fluctuation}. Based on this definition, the fluctuation theorems, such as the Jarzynski equality and the Crooks relation can be obtained straightforwardly \cite{jarzynski2011equalities, seifert2012stochastic, kurchan2000quantum, tasaki2000jarzynski, talkner2007fluctuation}. The TPM approach provides an operational way to measure the work in both isolated and open quantum systems, and the fluctuation theorems in the quantum regime have been tested experimentally using the TPM \cite{huber2008employing, batalhao2014experimental, an2015experimental}. In addition, it has been shown \cite{jarzynski2015quantum, zhu2016quantum, wang2017understanding, fei2018quantum, garcia2017quantum, arrais2018quantum, garcia2018semiclassical} that the definition of quantum fluctuating work based on the TPM obeys the quantum-classical correspondence principle, which provides some justification for this definition of quantum fluctuating work.

In spite of its success in the study of quantum thermodynamics, the TPM approach has its limitations which have been pointed out in recent studies \cite{perarnau2017no, lostaglio2018quantum, baumer2018fluctuating}. For states with quantum \emph{coherence}, the first projective energy measurement destroys the coherence, and affects the subsequent evolution of the system \cite{allahverdyan2014nonequilibrium, solinas2015full, hofer2016negative, solinas2016probing, solinas2017measurement, miller2017time, miller2018leggett, baumer2018fluctuating, perarnau2017no, lostaglio2018quantum, aberg2016fully, sampaio2018quantum, talkner2016aspects, francica2017role, xu2018effects, xu2019duality, wu2019experimentally}. 
Thus the averaged work is no longer equal to the difference of the internal energy (expectation value of the Hamiltonian) before and after the evolution. As a result, alternative definitions of quantum fluctuating work have been proposed for initial states with quantum coherence \cite{allahverdyan2005fluctuations, allahverdyan2014nonequilibrium, margenau1961correlation, solinas2015full, hofer2016negative, solinas2016probing, solinas2017measurement, nazarov2003full, miller2017time, sampaio2018quantum, talkner2016aspects}. Examples include that based on the full counting statistics (FCS) \cite{solinas2015full, hofer2016negative, solinas2016probing, solinas2017measurement, nazarov2003full}, and that based on the Margenau-Hill distribution (MH) \cite{allahverdyan2014nonequilibrium, margenau1961correlation}. The two definitions are related to the weak measurement that circumvents the invasive effect of the measurement disturbance on the statistics of work, and they are experimentally operational \cite{solinas2015full, hofer2016negative, solinas2016probing, solinas2017measurement, nazarov2003full, allahverdyan2014nonequilibrium, margenau1961correlation, johansen2007quantum, lundeen2012procedure}. When the initial state has no quantum coherence, the two definitions are equivalent to that based on the TPM \cite{baumer2018fluctuating, perarnau2017no, talkner2016aspects}. When the initial state has quantum coherence, the probabilities of work distributions are not positive-definite. In other words, they are quasi-probabilities \cite{baumer2018fluctuating, perarnau2017no, hofer2016negative, miller2017time}. For this reason, a no-go theorem for definitions of quantum work is proposed \cite{perarnau2017no, lostaglio2018quantum, baumer2018fluctuating, wu2019experimentally}. 

In the current study, we investigate the definitions of quantum fluctuating work based on the FCS and the MH from the perspective of quantum-classical correspondence. Inspired by Refs. \cite{jarzynski2015quantum, fei2018quantum} which studied the quantum-classical correspondence of quantum work distribution based on the TPM, we apply the same method of the phase space formulation of quantum mechanics \cite{wigner1997quantum, hillery1984distribution, polkovnikov2010phase, fei2018quantum}, and find that even in the presence of quantum coherence, both quantum work distributions converge to their classical counterpart in the limit of $\hbar \rightarrow 0$, where $\hbar$ is Planck's constant. In addition, we show that in comparison with the classical work, the two definitions of quantum fluctuating work lead to different quantum corrections 
\cite{wigner1997quantum, fei2018quantum}. 
Our results thus provide some justification for the validity of the definitions of quantum fluctuating work based on the FCS and the MH.

This paper is organized as follows. In Sec. \ref{Sec:FK}, we investigate the work characteristic functions based on the FCS and the MH by utilizing the quantum Feynman-Kac formula \cite{kac1949distributions, liu2012derivation, fei2018quantum} and the phase space formulation of quantum mechanics \cite{wigner1997quantum, hillery1984distribution, polkovnikov2010phase, fei2018quantum}, and give our main results. In Sec. \ref{Sec:HO}, we use an exactly solvable model of the linearly dragged harmonic oscillator to demonstrate our main results. In Sec. \ref{Sec:Conclusion}, we give some discussions and conclude our paper.

\section{\label{Sec:FK}Quantum-classical correspondence of the Feynman-Kac formula}

Our setup is an \emph{isolated} quantum system with an initial state described by a density matrix $\hat{\rho}(0)$. The system is driven by an external agent from the initial time $t = 0$ to the final time $t = \tau$. Accordingly, the Hamiltonian of the system is time-dependent $\hat{H}(t)$ that evolves from the initial time $t = 0$ to the final time $t = \tau$. 
The unitary evolution operator is
\beq
\hat{U}(\tau) = \overleftarrow{\mathcal{T}} \exp \sk{-\frac{\ii}{\hbar} \int_0^{\tau} \dd t \hat{H}(t)},
\label{Def:Unitary}
\eeq
where $\overleftarrow{\mathcal{T}}$ is the time-ordered operator. During the whole driving process, external work $W$ is exerted on the system. In this paper, we study the characteristic function of the work distribution $P(W)$:
\beq
\Phi(\eta) := \int_{-\infty}^{+\infty} \dd W P(W) \ee^{\ii \eta W}.
\label{Def:CF}
\eeq

If the initial state $\hat{\rho}(0)$ does not have quantum coherence, i.e., it is diagonal in the energy eigenbasis of $\hat{H}(0)$, we can adopt the TPM approach to define the quantum fluctuating work, and the characteristic function of work can be expressed as \cite{talkner2007fluctuation}:
\beq
\Phi_{\text{TPM}}(\eta) = \Tr \sk{\ee^{\ii \eta \hat{H}(\tau)} \hat{U}(\tau) \ee^{-\ii \eta \hat{H}(0)} \hat{\rho}(0) \hat{U}^{\dagger}(\tau)}.
\label{Def:TPM}
\eeq
If $\hat{\rho}(0)$ has quantum coherence, it does not commute with the initial Hamiltonian $\hat{H}(0)$, i.e., $\sk{\hat{\rho}(0),\hat{H}(0)} \neq 0$. In Ref. \cite{solinas2015full}, Solinas and Gasparinetti studied the quantum fluctuating work using the FCS. They gave the following definition of the characteristic function of work \cite{solinas2015full}:
\beq
\Phi_{\text{FCS}}(\eta) = \Tr \sk{\ee^{\ii \eta \hat{H}(\tau)} \hat{U}(\tau) \ee^{-\ii \frac{\eta}{2} \hat{H}(0)} \hat{\rho}(0) \ee^{-\ii \frac{\eta}{2} \hat{H}(0)} \hat{U}^{\dagger}(\tau)}.
\label{Def:FCS}
\eeq
The FCS is not the only way to characterize non-invasive measurements of work. In Ref. \cite{allahverdyan2014nonequilibrium}, Allahverdyan proposed another definition of the quantum fluctuating work based on the MH distribution for successive energy measurements \cite{allahverdyan2014nonequilibrium, margenau1961correlation, johansen2007quantum, lundeen2012procedure}:
\beq
\Phi_{\text{MH}}(\eta) = \Tr \sk{\ee^{\ii \eta \hat{H}(\tau)} \hat{U}(\tau) \rk{\ee^{-\ii \eta \hat{H}(0)} \star \hat{\rho}(0)} \hat{U}^{\dagger}(\tau)},
\label{Def:MH}
\eeq
where $\hat{A} \star \hat{B} := \half \sk{\hat{A} \hat{B} + \hat{B} \hat{A}}$.

Comparing Eqs. (\ref{Def:FCS}) and (\ref{Def:MH}) with \eqnref{Def:TPM}, we can see that $\Phi_{\text{FCS}}(\eta)$ and $\Phi_{\text{MH}}(\eta)$ are two types of symmetrization of $\Phi_{\text{TPM}}(\eta)$. They are introduced to deal with the situation when the initial state has quantum coherence (the initial state $\hat{\rho}(0)$ does not commute with the initial Hamiltonian $\hat{H}(0)$). 
While the first and the second moments of work, $\langle W \rangle$ and $\langle W^2 \rangle$, are the same for $\Phi_{\text{FCS}}(\eta)$ and $\Phi_{\text{MH}}(\eta)$, higher-order moments in general are different \cite{miller2018leggett}. It is worth mentioning that when the initial state has no coherence, the three definitions of quantum fluctuating work are equivalent.

\subsection{Phase space formulation of quantum Feynman-Kac formula}

In this section, we investigate the time evolution of the operators included in the trace of Eqs. (\ref{Def:FCS}) and (\ref{Def:MH}). For $\Phi_{\text{FCS}}(\eta)$ [\eqnref{Def:FCS}], let us define an operator $\hat{K}(t)$ as
\beq
\hat{K}(t) := \ee^{\ii \eta \hat{H}(t)} \hat{U}(t) \ee^{-\ii \frac{\eta}{2} \hat{H}(0)} \hat{\rho}(0) \ee^{-\ii \frac{\eta}{2} \hat{H}(0)} \hat{U}^{\dagger}(t).
\label{Def:K_FCS}
\eeq
It is easy to prove that $\hat{K}(t)$ satisfies the quantum Feynman-Kac formula introduced in Refs. \cite{liu2012derivation, fei2018quantum}:
\beq
\partial_t \hat{K}(t) = -\frac{\ii}{\hbar} \sk{\hat{H}(t),\hat{K}(t)} + \sk{\partial_t \ee^{\ii \eta \hat{H}(t)}} \ee^{-\ii \eta \hat{H}(t)} \hat{K}(t).
\label{Eqn:FK}
\eeq
The initial condition is
\beq
\hat{K}(0) = \ee^{\ii \frac{\eta}{2} \hat{H}(0)} \hat{\rho}(0) \ee^{-\ii \frac{\eta}{2} \hat{H}(0)}.
\label{Initial:FCS}
\eeq

We reformulate \eqnref{Eqn:FK} in the phase space (Weyl-Wigner) representation of quantum mechanics \cite{wigner1997quantum, hillery1984distribution, polkovnikov2010phase, fei2018quantum} as follows
\beqa
\partial_t K(z,t) &=& \frac{2}{\hbar} H(z,t) \sin \rk{\frac{\hbar}{2} \Lambda} K(z,t) \nonumber\\
& & + \Omega(z,t) \exp \rk{\frac{\ii \hbar}{2} \Lambda} K(z,t).
\label{Eqn:FK_WW}
\eeqa
This equation describes the time evolution of the function $K(z,t)$, which is the Weyl symbol of the operator $\hat{K}(t)$. Here $z = (x,p)$ represents a point in the phase space, and $x$ is the position and $p$ is the momentum of the particle. In \eqnref{Eqn:FK_WW}, $H(z,t)$ is the Weyl symbol of the Hamiltonian $\hat{H}(t)$. In this paper, for simplicity, we study a system described by the following Hamiltonian
\beq
\hat{H}(t) = \frac{\hat{p}^2}{2 m} + V(\hat{x},t),
\label{Hamiltonian}
\eeq
where $m$ is the mass of the single particle, and $V$ is the potential. The symplectic operator $\Lambda$ \cite{hillery1984distribution} is
\beq
\Lambda = \overleftarrow{\partial_x} \overrightarrow{\partial_p} - \overleftarrow{\partial_p} \overrightarrow{\partial_x},
\label{Def:Lambda}
\eeq
and the arrows denote the directions the partial derivatives act upon. In \eqnref{Eqn:FK_WW},
\beq
\Omega(z,t) = \bk{\partial_t \ee^{\ii \eta \hat{H}(t)}}_w \exp \rk{\frac{\ii \hbar}{2} \Lambda} \bk{\ee^{-\ii \eta \hat{H}(t)}}_w,
\label{Def:Omega}
\eeq
where $\{ \cdot \}_w$ denotes the Weyl symbol of the operator included in the bracket, and $\exp \rk{\frac{\ii \hbar}{2} \Lambda}$ denotes the star product \cite{hillery1984distribution}. Then according to \eqnref{Def:FCS}, the characteristic function of quantum work can be calculated as
\beq
\Phi(\eta) = \int_{-\infty}^{+\infty} \dd z K(z,\tau),
\label{Eqn:CF}
\eeq
where the integral over $\dd z$ denotes the integral over the whole phase space.

Having introduced the phase space formulation of quantum Feynman-Kac formula, in the following, we study the quantum work statistics and its relation to the classical work statistics.

\subsection{Quantum work statistics and its classical counterpart}

As pointed out by Wigner in 1932 \cite{wigner1997quantum}, the phase space representation has advantages in studying the quantum-classical correspondence and calculating the quantum corrections on thermodynamic variables in powers of $\hbar$. In this subsection, we calculate the work statistics by solving \eqnref{Eqn:FK_WW}. Let us recall that when studying the Weyl symbol of the exponential of the Hamiltonian $\ee^{-\ii \eta \hat{H}(t)}$, for example, the density matrix of the thermal equilibrium state $\hat{\rho}^{eq}(0) \propto \ee^{-\beta \hat{H}(0)}$ ($\beta$ is the inverse temperature), Wigner found that it can be expanded as \cite{wigner1997quantum}
\beq
\bk{\ee^{-\ii \eta \hat{H}(t)}}_w = \ee^{-\ii \eta H(z,t)} \sk{1 + (\ii \hbar)^2 f(\ii \eta,z,t) + \mathcal{O} (\hbar^4)},
\label{Wigner expansion}
\eeq
where
\beq
f(\ii \eta,z,t) = \frac{(\ii \eta)^2}{8 m} \sk{\partial_x^2 V - \frac{\ii \eta}{3} \rk{\partial_x V}^2 - \frac{\ii \eta}{3 m} p^2 \partial_x^2 V}.
\label{Def:f}
\eeq
Please note that on the r.h.s. of \eqnref{Wigner expansion}, there are no terms proportional to odd orders of $\hbar$, and this feature is important for our later analysis. By substituting Eqs. (\ref{Def:Omega}) and (\ref{Wigner expansion}) into  \eqnref{Eqn:FK_WW} and expanding \eqnref{Eqn:FK_WW} in powers of $\hbar$, we obtain
\beqa
\partial_t K = & & H \Lambda K + \ii \eta \partial_t H K\nonumber\\
& & + \frac{\ii \hbar}{2} \sk{(\ii \eta)^2 (H \Lambda \partial_t H) + \ii \eta \partial_t H \Lambda} K + \mathcal{O} (\hbar^2).
\label{Eqn:FK_WW2}
\eeqa
By expanding $K(z,t)$ in powers of $\hbar$:
\beq
K(z,t) = K^{(0)}(z,t) + (\ii \hbar) K^{(1)}(z,t) + (\ii \hbar)^2 K^{(2)}(z,t) + \cdots,
\label{K expansion}
\eeq
and identifying terms of the same orders of $\hbar$ in \eqnref{Eqn:FK_WW2}, we obtain
\beqa
\partial_t K^{(0)} = & & H \Lambda K^{(0)} + \ii \eta \partial_t H K^{(0)};\nonumber\\
\partial_t K^{(1)} = & & H \Lambda K^{(1)} + \ii \eta \partial_t H K^{(1)}\nonumber\\
& & + \half \sk{(\ii \eta)^2 (H \Lambda \partial_t H) + \ii \eta \partial_t H \Lambda} K^{(0)};\nonumber\\
\cdots
\label{Eqn:FK_hbars}
\eeqa
The initial conditions [\eqnref{Initial:FCS}] in the phase space representation are (see Appendix \ref{A:Initial})
\beqa
K^{(0)}(z,0) &=& P^{(0)}(z,0);\nonumber\\
K^{(1)}(z,0) &=& \frac{\ii \eta}{2} H(z,0) \Lambda P^{(0)}(z,0);\nonumber\\
&\cdots& 
\label{Initial:FCS_WW}
\eeqa
where $P^{(0)}(z,0)$ denotes the zeroth-order term of the Weyl symbol of the initial density matrix $\hat{\rho}(0)$, and it will be defined later [see \eqnref{States: Even hbar}]. We regard $P^{(0)}(z,0)$ as the classical counterpart of $\hat{\rho}(0)$: 
\beq
P_{\text{Classical}}(z,0) := P^{(0)}(z,0).
\label{QCC:initial state}
\eeq
It can be checked \cite{wigner1997quantum} that for the quantum thermal equilibrium state, $P^{(0)}(z,0)$ is exactly the thermal equilibrium state of the classical Hamiltonian $H(z,0)$.

Please note that not all quantum states have the classical counterparts in general. For example, the Fock states in quantum optics or the spin states in spin systems do not have well-defined classical counterparts. However, for systems described by \eqnref{Hamiltonian}, the thermal equilibrium state \cite{wigner1997quantum} and those evolved from the thermal equilibrium state have well-defined classical counterparts in the phase space. For simplicity, we focus on those states evolved from the thermal equilibrium state in the following discussions. The Wigner functions of these states have the following form in powers of $\hbar$:
\beq
P(z,0) = P^{(0)}(z,0) + (\ii \hbar)^2 P^{(2)}(z,0) + \mathcal{O}(\hbar^4).
\label{States: Even hbar}
\eeq
That is to say, their Wigner functions do not contain terms proportional to odd orders of $\hbar$. The reason is given below \eqnref{Eqn:EOM_P}. These states may have quantum coherence, and meanwhile they have well-defined classical counterparts.

The equation of motion of $K^{(0)}(z,t)$ [\eqnref{Eqn:FK_hbars}] is the classical Liouville equation plus the additional term $\ii \eta \partial_t H K^{(0)}$. By utilizing the classical Feynman-Kac formula \cite{kac1949distributions}, we obtain  the solution, which is the conditional expectation of the classical work functional:
\beq
K^{(0)}(z,\tau) = \av{\ee^{\ii \eta W \sk{z(t)}} \delta \rk{z-z(\tau)}}_{P^{(0)}(z,0)},
\label{Sol:K0}
\eeq
where
\beq
W \sk{z(t)} = \int_0^\tau \partial_t H \sk{z(t),t} \dd t
\label{Def:Classical work}
\eeq
is the classical work functional of the stochastic trajectories \cite{jarzynski1997nonequilibrium, sekimoto1998langevin, sekimoto2010stochastic}, and the brackets in \eqnref{Sol:K0} denotes the average over all classical trajectories in the phase space starting from the initial distribution $P^{(0)}(z,0)$. Since it is an isolated system, the trajectories satisfy Newton's equation.

If the characteristic function of work can be expanded as
\beq
\Phi(\eta) = \Phi^{(0)}(\eta) + (\ii \hbar) \Phi^{(1)}(\eta) + (\ii \hbar)^2 \Phi^{(2)}(\eta) + \cdots,
\label{Phi expansion}
\eeq
then according to \eqnref{Eqn:CF}, the zeroth order of the characteristic function of quantum work is
\beq
\Phi^{(0)}(\eta) = \av{\ee^{\ii \eta \int_0^\tau \partial_t H \sk{z(t),t} \dd t}}_{P^{(0)}(z,0)},
\label{Sol:Phi0}
\eeq
where the r.h.s. of \eqnref{Sol:Phi0} is exactly the classical work characteristic function $\Phi_{\text{Classical}}(\eta)$ with the initial distribution $P_{\text{Classical}}(z,0)$, i.e.,
\beq
\Phi^{(0)}(\eta) = \Phi_{\text{Classical}}(\eta) \equiv \av{\ee^{\ii \eta \int_0^\tau \partial_t H \sk{z(t),t} \dd t}}_{P_{\text{Classical}}(z,0)}.
\label{QCC_FK}
\eeq

By comparing the first two equations of \eqnref{Eqn:FK_hbars}, we obtain
\beq
K^{(1)} = \frac{\ii \eta}{2} H \Lambda K^{(0)}.
\label{Eqn:K1K0}
\eeq
Using the property of the symplectic operator \cite{hillery1984distribution}, we find that the first-order correction of $\Phi(\eta)$ vanishes:
\beq
\Phi^{(1)}(\eta) = \int_{-\infty}^{+\infty} \dd z K^{(1)}(z,\tau) \equiv 0.
\label{Sol:Phi1}
\eeq
Similar to the results of Ref. \cite{wigner1997quantum} [\eqnref{Wigner expansion}], the corrections of $\Phi(\eta)$ proportional to odd orders of $\hbar$ vanish.

Up to now, we have discussed the characteristic function based on the FCS. For that based on the MH distribution, the definition of the operator $\hat{K}(t)$ is different. According to \eqnref{Def:MH},
\beq
\hat{K}(t) := \ee^{\ii \eta \hat{H}(t)} \hat{U}(t) \rk{\ee^{-\ii \eta \hat{H}(0)} \star \hat{\rho}(0)} \hat{U}^{\dagger}(t).
\label{Def:K_MH}
\eeq
All the equations listed above [from \eqnref{Eqn:FK} to \eqnref{Sol:Phi1}] are equally applicable to the study of the characteristic function based on the MH except that the initial condition \eqnref{Initial:FCS} should be replaced by
\beq
\hat{K}(0) = \half \sk{\hat{\rho}(0) + \ee^{\ii \eta \hat{H}(0)} \hat{\rho}(0) \ee^{-\ii \eta \hat{H}(0)}}.
\label{Initial:MH}
\eeq
Please note that the initial conditions of $K^{(0)}(z,t)$ and $K^{(1)}(z,t)$ are still the same as \eqnref{Initial:FCS_WW} (see Appendix \ref{A:Initial}). Accordingly, the classical counterparts [zeroth order, \eqnref{Sol:Phi0}] and the corrections proportional to $\hbar$ [\eqnref{Sol:Phi1}] are the same for $\Phi_{\text{FCS}}(\eta)$ and $\Phi_{\text{MH}}(\eta)$. Nevertheless, due to the difference between the initial conditions [\eqnref{Initial:FCS} and \eqnref{Initial:MH}], the corrections proportional to the second and higher even orders of $\hbar$ are different. In the following section, we will use an exactly solvable model to demonstrate our main results.

\section{\label{Sec:HO}A case study: Linearly dragged harmonic oscillator}

In the preceding section, we focus our attention on the general cases whose Hamiltonian is in the form of \eqnref{Hamiltonian}, and the initial quantum states are those evolved from the thermal equilibrium state [\eqnref{States: Even hbar}]. In this section, as an illustration, we study a linearly dragged harmonic oscillator. The Hamiltonian is
\beq
\hat{H}(t) = \frac{\hat{p}^2}{2 m} + \half m \omega^2 \rk{\hat{x} - u t}^2,
\label{Hamiltonian:HO}
\eeq
where $\omega$ is the trapping frequency of the harmonic potential, and $u$ is the speed of the shifting of the potential center. At time $t = -\tau'$ ($\tau' > 0$), the system is prepared in the thermal equilibrium state $\hat{\rho}(-\tau') = \frac{1}{Z_q} \ee^{-\beta \hat{H}(-\tau')}$, where $Z_q$ is the quantum partition function \cite{talkner2008statistics, deffner2008nonequilibrium}:
\beq
Z_q = \frac{1}{2 \sinh \frac{\beta \hbar \omega}{2}}.
\label{Partition function:Quantum}
\eeq
The corresponding classical state is the equilibrium state of the classical Hamiltonian \cite{wigner1997quantum} $P^{(0)}(z,-\tau') = \frac{1}{Z_c} \ee^{-\beta H(z,-\tau')}$, where $Z_c$ is the classical partition function:
\beq
Z_c = \frac{2 \pi}{\beta \omega}.
\label{Partition function:Classical}
\eeq
The Wigner function $P(z,t)$ of the density matrix $\hat{\rho}(t)$ satisfies the following equation of motion 
\cite{wigner1997quantum, hillery1984distribution}:
\beq
\partial_t P(z,t) = H(z,t) \Lambda P(z,t) + \frac{(\ii \hbar)^2}{24} \partial_x^3 V \partial_p^3 P(z,t) + \cdots,
\label{Eqn:EOM_P}
\eeq
where terms proportional to odd orders of $\hbar$ vanish on the r.h.s. of \eqnref{Eqn:EOM_P} \cite{wigner1997quantum, hillery1984distribution}. Since $P(z,-\tau')$ does not contain terms proportional to odd orders of $\hbar$ [see \eqnref{Wigner expansion}], from \eqnref{Eqn:EOM_P}, we conclude that $P(z,t)$ does not contain terms proportional to odd orders of $\hbar$ 
[see \eqnref{States: Even hbar}]. In our model [\eqnref{Hamiltonian:HO}], $\partial_x^3 V = 0$. Thus
\beq
\partial_t P(z,t) = H(z,t) \Lambda P(z,t),
\label{Eqn:EOM_PHO}
\eeq
which is the classical Liouville equation.

The system evolves under the governing of $\hat{H}(t)$ from $t = -\tau'$ to $t = 0$, and we regard the final state $\hat{\rho}(0)$ of the first evolution as the initial state of the second evolution. The final time of the second evolution is $t = \tau$, and the final state is $\hat{\rho}(\tau)$. We care about the quantum work distribution of the second evolution only (from $t = 0$ to $t = \tau$). Please note that as long as the first evolution (from $t = -\tau'$ to $t = 0$) is not quantum adiabatic, the initial state of the second evolution $\hat{\rho}(0)$ has quantum coherence, i.e., $\sk{\hat{\rho}(0),\hat{H}(0)} \neq 0$.

According to \eqnref{Eqn:EOM_P}, the equation of motion of $P^{(0)}(z,t)$ is the classical Liouville equation with the initial condition $P^{(0)}(z,-\tau') = \frac{1}{Z_c} \ee^{-\beta H(z,-\tau')}$. Hence, the evolution of $P^{(0)}(z,t)$ is purely classical under the governing of the classical Hamiltonian $H(z,t)$, i.e., the whole quantum driving process has a perfect classical counterpart which evolves from $P^{(0)}(z,-\tau')$ to $P^{(0)}(z,0)$ and finally to $P^{(0)}(z,\tau)$ under the governing of the classical Hamiltonian $H(z,t)$. Thus we can regard $P^{(0)}(z,t)$ as the classical counterpart of $\hat{\rho}(t)$. The classical characteristic function of work of the second evolution (from $t = 0$ to $t = \tau$) $\Phi_{\text{Classical}}(\eta)$ can be obtained from Ref. \cite{pan2018quantifying}:
\begin{widetext}
\beq
\Phi_{\text{Classical}}(\eta) = \exp \bk{m u^2 \sk{\ii \eta \rk{\cos \omega \tau' - \cos \omega (\tau' + \tau)} - \frac{\eta^2}{\beta} \rk{1 - \cos \omega \tau}}},
\label{Sol:PhiClassical_HO}
\eeq
\end{widetext}

From Refs. \cite{wigner1997quantum, hillery1984distribution, polkovnikov2010phase}, we know that 
\begin{widetext}
\beq
P(z,-\tau') = \frac{1}{\pi \hbar \coth \frac{\beta \hbar \omega}{2}} \exp \bk{-\frac{p^2}{m \hbar \omega \coth \frac{\beta \hbar \omega}{2}} - \frac{\sk{x + u \tau'}^2}{\frac{\hbar}{m \omega} \coth \frac{\beta \hbar \omega}{2}}}.
\label{WF:Pz-t1}
\eeq
\end{widetext}
By solving the Liouville equation [\eqnref{Eqn:EOM_PHO}], we obtain \cite{pan2018quantifying}
\begin{widetext}
\beq
P(z,0) = \frac{1}{\pi \hbar \coth \frac{\beta \hbar \omega}{2}} \exp \bk{-\frac{\sk{p - m u \rk{1 - \cos \omega \tau'}}^2}{m \hbar \omega \coth \frac{\beta \hbar \omega}{2}} - \frac{\sk{x + \frac{u}{\omega} \sin \omega \tau'}^2}{\frac{\hbar}{m \omega} \coth \frac{\beta \hbar \omega}{2}}}.
\label{WF:Pz0}
\eeq
\end{widetext}
We would like to emphasize that the initial state of the second evolution [\eqnref{WF:Pz0}] has quantum coherence. The corresponding classical state is
\begin{widetext}
\beq
P^{(0)}(z,0) = \frac{\beta \omega}{2 \pi} \exp \bk{-\frac{\beta}{2 m} \sk{p - m u \rk{1 - \cos \omega \tau'}}^2 - \half \beta m \omega^2 \sk{x + \frac{u}{\omega} \sin \omega \tau'}^2} \equiv P_{\text{Classical}}(z,0).
\label{State:0}
\eeq
\end{widetext}

Now that we have obtained the quantum-classical correspondence of the initial state of the second evolution, we will analytically calculate the zeroth orders of the work characteristic functions $\Phi_{\text{FCS}}(\eta)$ and $\Phi_{\text{MH}}(\eta)$. Let us consider $\Phi_{\text{FCS}}(\eta)$ first. By separating $\hat{K}(t)$ [\eqnref{Def:K_FCS}] into two parts:
\beq
\hat{K}(t) = \ee^{\ii \eta \hat{H}(t)} \hat{M}(t),
\label{Def:KM_FCS}
\eeq
where
\beq
\hat{M}(t) := \hat{U}(t) \ee^{-\ii \frac{\eta}{2} \hat{H}(0)} \hat{\rho}(0) \ee^{-\ii \frac{\eta}{2} \hat{H}(0)} \hat{U}^{\dagger}(t),
\label{Def:M_FCS}
\eeq
it is easy to prove that the equation of motion of $\hat{M}(t)$ is the quantum Liouville-von Neumann equation:
\beq
\partial_t \hat{M}(t) = -\frac{\ii}{\hbar} \sk{\hat{H}(t),\hat{M}(t)},
\label{Eqn:EOM_M}
\eeq
with the initial condition
\beq
\hat{M}(0) = \ee^{-\ii \frac{\eta}{2} \hat{H}(0)} \hat{\rho}(0) \ee^{-\ii \frac{\eta}{2} \hat{H}(0)}.
\label{Initial:M_FCS}
\eeq
Similarly, we reformulate them in the phase space representation. We define the Weyl symbol of $\hat{M}(t)$ as $\bk{\hat{M}(t)}_w := M(z,t)$. Due to the peculiarity of the harmonic oscillator, \eqnref{Eqn:EOM_M} corresponds to the classical Liouville equation:
\beq
\partial_t M(z,t) = H(z,t) \Lambda M(z,t).
\label{Eqn:EOM_M_WW}
\eeq
Using Eqs. (\ref{Wigner expansion}), (\ref{WF:Pz0}), (\ref{State:0}) and the conclusion of Appendix \ref{A:Initial}, we find that $M(z,0)$ can be expanded in even orders of $\hbar$:
\beq
M(z,0) = \ee^{-\ii \eta H(z,0)} P^{(0)}(z,0) + (\ii \hbar)^2 M^{(2)}(z,0) + \cdots.
\label{Mz0 expansion}
\eeq

The characteristic function can be calculated as
\beq
\Phi(\eta) = \int_{-\infty}^{+\infty} \dd z \bk{\ee^{\ii \eta \hat{H}(\tau)}}_w M(z,\tau).
\label{Eqn:CF_WW}
\eeq
Thus the zeroth order is
\beq
\Phi^{(0)}(\eta) = \int_{-\infty}^{+\infty} \dd z \ \ee^{\ii \eta H(z,\tau)} M^{(0)}(z,\tau),
\label{Eqn:CF_WW0}
\eeq
and the first order vanishes because neither $\bk{\ee^{\ii \eta \hat{H}(\tau)}}_w$ nor $M(z,\tau)$ contains terms proportional to odd orders of $\hbar$:
\beq
\Phi^{(1)}(\eta) = 0.
\label{Eqn:CF_WW1}
\eeq
By solving the classical Liouville equation, we can obtain $M^{(0)}(z,\tau)$ that evolves from $M^{(0)}(z,0) = \ee^{-\ii \eta H(z,0)} P^{(0)}(z,0)$. After some algebra (see Appendix \ref{A:HO0}), we obtain
\begin{widetext}
\beq
\Phi^{(0)}(\eta) = \exp \bk{m u^2 \sk{\ii \eta \rk{\cos \omega \tau' - \cos \omega (\tau' + \tau)} - \frac{\eta^2}{\beta} \rk{1 - \cos \omega \tau}}},
\label{Sol:Phi0_HO}
\eeq
\end{widetext}
which is identical to the classical characteristic function of the work $\Phi_{\text{Classical}}(\eta)$ [\eqnref{Sol:PhiClassical_HO}] with the initial state $P_{\text{Classical}}(z,0)$ \cite{pan2018quantifying}:
\beq
\Phi^{(0)}(\eta) = \Phi_{\text{Classical}}(\eta).
\label{QCC_HO}
\eeq
Thus we have obtained the quantum-classical correspondence of work distributions based on the FCS.

For the MH, according to \eqnref{Def:MH}, the operator $\hat{M}(t)$ should be replaced by
\beq
\hat{M}(t) := \hat{U}(t) \rk{\ee^{-\ii \eta \hat{H}(0)} \star \hat{\rho}(0)} \hat{U}^{\dagger}(t),
\label{Def:M_MH}
\eeq
and the initial condition
\beq
\hat{M}(0) = \ee^{-\ii \eta \hat{H}(0)} \star \hat{\rho}(0).
\label{Initial:M_MH}
\eeq
Both Eqs. (\ref{Eqn:EOM_M}) and (\ref{Eqn:EOM_M_WW}) also apply to $\Phi_{\text{MH}}(\eta)$. We can prove that $M(z,0)$ contains only terms proportional to even orders of $\hbar$, and the zeroth order is the same as that based on the FCS (see Appendix \ref{A:Initial}). By using \eqnref{Eqn:CF_WW}, we find that
\beqa
\Phi^{(0)}_{\text{FCS}}(\eta) &=& \Phi^{(0)}_{\text{MH}}(\eta) = \Phi_{\text{Classical}}(\eta),
\label{FCSMH:Phi0}\\
\Phi^{(1)}_{\text{FCS}}(\eta) &=& \Phi^{(1)}_{\text{MH}}(\eta) = 0.
\label{FCSMH:Phi1}
\eeqa
Since $M^{(2)}_{\text{FCS}}(z,0) \neq M^{(2)}_{\text{MH}}(z,0)$, we obtain
\beq
\Phi^{(2)}_{\text{FCS}}(\eta) \neq \Phi^{(2)}_{\text{MH}}(\eta).
\label{FCSMH:Phi2}
\eeq

\begin{figure*}[t]
	\includegraphics[scale=0.45,angle=0]{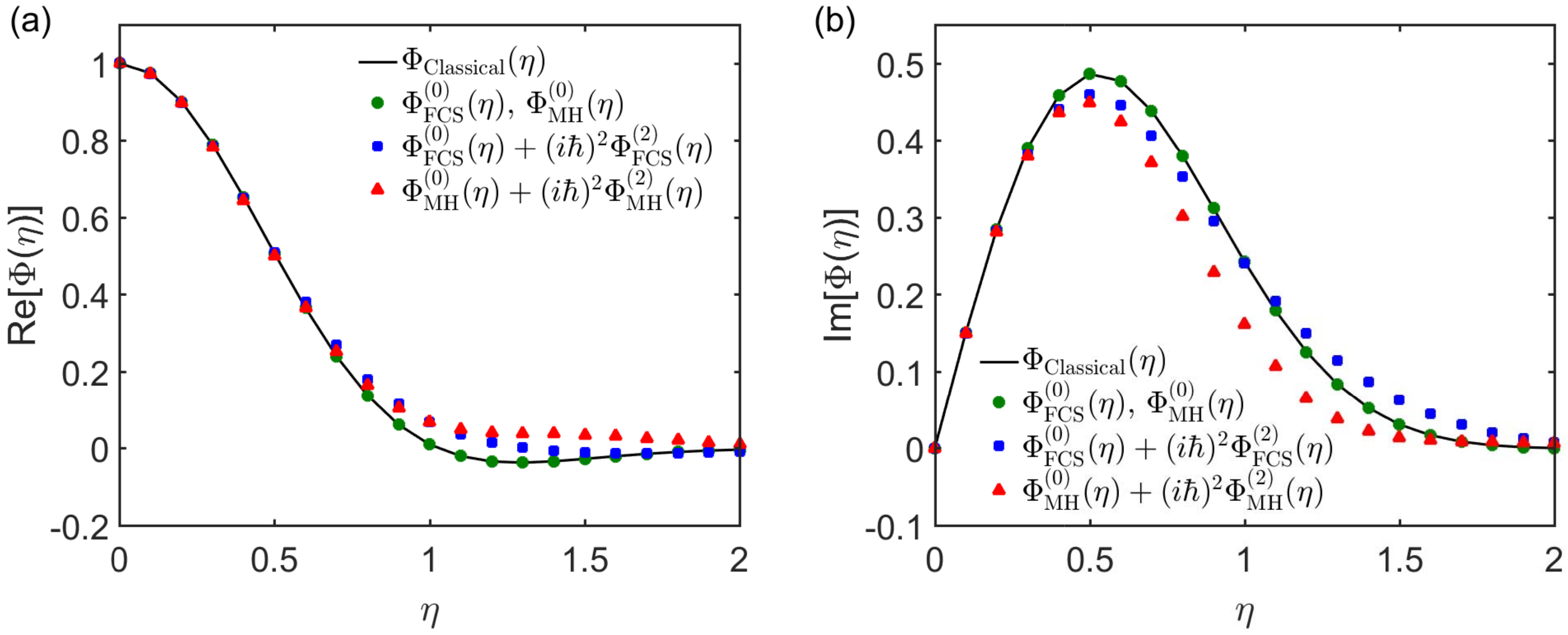}
	\caption{(Color online) Different orders of characteristic functions of work distributions based on the FCS and MH. (a) Real part. (b) Imaginary part. Here $\beta = u = m = \omega = \hbar = 1$, $\tau' = 1$, $\tau = 2$. In each figure, there are three sets of points and one curve. The black curve represents the classical characteristic function $\Phi_{\text{Classical}}(\eta)$ [\eqnref{Sol:PhiClassical_HO}]. The green circles represent the zeroth order ($\Phi^{(0)}_{\text{FCS}}(\eta)$ and $\Phi^{(0)}_{\text{MH}}(\eta)$) [\eqnref{FCSMH:Phi0}]. The blue squares represent the zeroth plus the second orders for the work definition based on the FCS. The red triangles represent the zeroth plus the second orders for the work definition based on the MH.
	}
	\label{Fig:1}
\end{figure*}

The detailed calculations of $\Phi^{(2)}_{\text{FCS}}(\eta)$ and $\Phi^{(2)}_{\text{MH}}(\eta)$ are shown in Appendix \ref{A:HO0}. In \figref{}{Fig:1}, we plot the real and the imaginary parts of $\Phi_{\text{Classical}}(\eta)$, $\Phi^{(0)}_{\text{FCS}}(\eta)$, $\Phi^{(0)}_{\text{MH}}(\eta)$, $\Phi^{(0)}_{\text{FCS}}(\eta)+(\ii \hbar)^2 \Phi^{(2)}_{\text{FCS}}(\eta)$, and $\Phi^{(0)}_{\text{MH}}(\eta)+(\ii \hbar)^2 \Phi^{(2)}_{\text{MH}}(\eta)$ of the linearly dragged harmonic oscillator. It is easy to see the quantum-classical correspondence and the differences between $\Phi_{\text{FCS}}(\eta)$ and $\Phi_{\text{MH}}(\eta)$.


\section{\label{Sec:Conclusion}Discussion and conclusion}

\subsection{Generalized Jarzynski equalities for arbitrary initial states with quantum coherence}

Based on the FCS and the MH, one can define the quantum fluctuating work for quantum systems with initial coherence. One of the applications of these definitions of quantum fluctuating work is to study the Jarzynski equality for arbitrary initial states. In Ref. \cite{gong2015jarzynski}, the Jarzynski equality \cite{jarzynski1997nonequilibrium, seifert2012stochastic, jarzynski2011equalities} is extended to arbitrary initial states in the \emph{classical} regime. If the initial distribution in the phase space is $p(x,p,0)$, then
\beq
\av{\ee^{-\beta \rk{W - \Delta F}}}_{p(x,p,0)} = \iint \dd x \dd p \ p^R(x,-p,\tau) \frac{p(x,p,0)}{p^{eq}(x,p,0)},
\label{Eqn:GJE_Gong}
\eeq
where $p^{eq}(x,p,0)$ is the equilibrium distribution at the initial time $t = 0$, and $p^R(x,-p,\tau)$ is the final distribution of the corresponding time-reversal process \cite{crooks1999entropy, jarzynski2011equalities, seifert2012stochastic} whose initial state is chosen to be the equilibrium state.

In the quantum regime for isolated quantum systems, if the initial state $\hat{\rho}(0)$ does not have coherence, \eqnref{Eqn:GJE_Gong} can be straightforwardly generalized to the quantum version by adopting the TPM approach:
\beqa
\av{\ee^{-\beta \rk{W - \Delta F}}}_{\hat{\rho}(0)} &=& \sum_n \rho_{nn}^R(\tau) \frac{\rho_{nn}(0)}{\rho_{nn}^{eq}(0)} \nonumber\\
&=& \Tr \sk{\hat{\rho}(0) \hat{\rho}^R(\tau) \rk{\hat{\rho}^{eq}(0)}^{-1}},
\label{Eqn:GJE_QnoCoherence}
\eeqa
where $\hat{\rho}^{eq}(0)$ is the equilibrium density matrix at the initial time $t = 0$, and $\hat{\rho}^R(\tau)$ is the final density matrix of the corresponding time-reversal process whose initial state is chosen to be the equilibrium density matrix. The subscript $nn$ depicts the $n$-th diagonal matrix element.

If the initial state $\hat{\rho}(0)$ has coherence, then according to the work definition based on the FCS [\eqnref{Def:FCS}], we can similarly generalize the Jarzynski equality to arbitrary initial states with quantum coherence:
\beq
\av{\ee^{-\beta \rk{W - \Delta F}}}_{\hat{\rho}(0)} = \sum_{m,n} \rho_{mn}^R(\tau) \frac{\rho_{nm}(0)}{\sqrt{\rho_{mm}^{eq}(0) \rho_{nn}^{eq}(0)}},
\label{Eqn:GJE_QwithCoherence_FCS}
\eeq
where the subscript $mn$ depicts the matrix element in the $m$-th row and the $n$-th column. For the work definition based on the MH [\eqnref{Def:MH}], the generalized Jarzynski equality for arbitrary initial states with quantum coherence is \cite{allahverdyan2014nonequilibrium}
\beq
\av{\ee^{-\beta \rk{W - \Delta F}}}_{\hat{\rho}(0)} = \text{Re} \ \Tr \sk{\hat{\rho}(0) \hat{\rho}^R(\tau) \rk{\hat{\rho}^{eq}(0)}^{-1}}.
\label{Eqn:GJE_QwithCoherence_MH}
\eeq
Thus by adopting work definitions based on the FCS and the MH, 
we extend Jarzynski equality from equilibrium initial states to arbitrary initial states with quantum coherence.

\subsection{Effect of the first projective energy measurement}

Quantum coherence of the initial states is the main concern of our current study. We may ask a question: does the quantum-classical correspondence of work distributions still hold if we make a projective energy measurement at the initial time? For initial states without coherence, the answer is yes because the states remain unchanged after the first projective measurement.

For the initial states with quantum coherence, in the energy basis, the off-diagonal elements will be destroyed by the measurement. Initially, $\sk{\hat{\rho}(0),\hat{H}(0)} \neq 0$. The initial Wigner function of $\hat{\rho}(0)$ is $P(z,0)$. After the projective energy measurement, $\hat{\rho}(0)$ becomes $\hat{\tilde{\rho}}(0)$, and the Wigner function becomes
\beq
\tilde{P}(r,0) = \frac{1}{2 \pi} \int_0^{2 \pi} \dd \theta' P(r,\theta',0),
\label{WF measurement}
\eeq
where $P(r,\theta,0)$ and $\tilde{P}(r,\theta,0)$ are expressed in the polar coordinate, corresponding to $P(z,0)$ and $\tilde{P}(z,0)$ respectively. The derivation of \eqnref{WF measurement} is given in Appendix \ref{A:Measurement}. As we can see, $\tilde{P}$ is the average of $P$ over the angular coordinate $\theta$, thus $\tilde{P}$ is independent of $\theta$. Since $P(r,\theta,0)$ [\eqnref{WF:Pz0}] and $P^{(0)}(r,\theta,0)$ [\eqnref{State:0}] are $\theta$-dependent, we obtain
\beqa
\tilde{P}(z,0) &\neq& P(z,0),
\label{WF measurement 2}\\
\tilde{P}^{(0)}(z,0) &\neq& P^{(0)}(z,0).
\label{WF0 measurement}
\eeqa
The change of the initial state due to the first projective energy measurement will influence the work statistics. Thus after the first projective energy measurement, both the initial state and the work statistics will change in all orders of $\hbar$, including the zeroth order.


\subsection{Wigner function and quasi-probability}

In quantum mechanics, a point in the phase space is not a proper way to describe the state of the system due to the uncertainty principle. 
Nevertheless, in order to compare with its classical counterpart, physicists introduced a quasi-probability known as the Wigner function. But this quasi-probability may be negative. The leading order of the Wigner function is equal to the classical distribution in the phase space \cite{wigner1997quantum}, which provides some justification for the concept of the Wigner function as a quasi-probability. Similarly, in quantum thermodynamics, when dealing with quantum systems with initial coherence, it is improper to define quantum fluctuating work along individual stochastic trajectories in the phase space. Nevertheless, 
physicists introduced the quantum fluctuating work based on the FCS or the MH, but the work distribution may be negative (quasi-probability). We find that the leading order of the work distribution is equal to the classical counterpart, which implies the quantum-classical correspondence of work distributions, and provides some justification for definitions of quantum fluctuating work based on the FCS and the MH.

In summary, in recent years, definitions of quantum fluctuating work based on the FCS and the MH have been proposed for initial states with quantum coherence, and have attracted a lot of attention. But these definitions seem ad hoc. In this article, we study the quantum-classical correspondence of work distributions based on the two definitions. 
Firstly, for the general cases, we prove that both definitions satisfy the quantum Feynman-Kac formula \cite{kac1949distributions, liu2012derivation, fei2018quantum}. Using the method of the phase space formulation of quantum mechanics \cite{wigner1997quantum, hillery1984distribution, polkovnikov2010phase, fei2018quantum} and $\hbar$ expansion, we prove that the leading order of both work distributions corresponds to its classical counterpart. Corrections proportional to odd orders of $\hbar$ vanish, and corrections proportional to $\hbar^2$ are different. Then, as an exactly solvable example, we calculate the leading order and the second order of work distributions of a linearly dragged harmonic oscillator. We use this example to demonstrate the quantum-classical correspondence of work distributions and the quantum corrections. In addition, we discuss the generalized Jarzynski equalities for arbitrary initial states with quantum coherence based on the FCS and the MH. 
Our work is an extension of previous work for the definition based on the TPM \cite{jarzynski2015quantum, zhu2016quantum, wang2017understanding, fei2018quantum, garcia2017quantum, arrais2018quantum, garcia2018semiclassical}, and provides some justification for the validity of the definitions of quantum fluctuating work based on the FCS and the MH.

\begin{acknowledgments}
The authors thank Professor Christopher Jarzynski for helpful discussions. H. T. Quan acknowledges support from the National Science Foundation of China under Grants No. 11775001, No. 11534002, and No. 11825501.
\end{acknowledgments}

\appendix

\setcounter{figure}{0}
\renewcommand{\thefigure}{A\arabic{figure}}

\section{\label{A:Initial}Derivation of Eqs. (\ref{Initial:FCS_WW}) and (\ref{Mz0 expansion})}

Firstly, let us derive \eqnref{Initial:FCS_WW}. From the initial condition of $\hat{K}(t)$ based on $\Phi_{\text{FCS}}(\eta)$ [\eqnref{Initial:FCS}], we obtain the Weyl symbol of the operator $\hat{K}(0)$:
\begin{widetext}
\beq
K(z,0) = \sk{\bk{\ee^{\ii \frac{\eta}{2} \hat{H}(0)}}_w \exp \rk{\frac{\ii \hbar}{2} \Lambda} P(z,0)} \exp \rk{\frac{\ii \hbar}{2} \Lambda} \bk{\ee^{-\ii \frac{\eta}{2} \hat{H}(0)}}_w.
\label{Kz0_FCS}
\eeq
\end{widetext}
We expand $K(z,0)$ in powers of $\hbar$, and we obtain the zeroth order when we take the zeroth order of each term in the Weyl symbol:
\beqa
K^{(0)}(z,0) &=& \ee^{\ii \frac{\eta}{2} H(z,0)} P^{(0)}(z,0) \ee^{-\ii \frac{\eta}{2} H(z,0)} \nonumber\\
&=& P^{(0)}(z,0).
\label{Kz0_0_FCS}
\eeqa
Since the initial state $P(z,0)$ only contains terms proportional to even orders of $\hbar$ [\eqnref{States: Even hbar}], the first order of $K(z,0)$ originating from the two star products $\exp \rk{\frac{\ii \hbar}{2} \Lambda}$ is
\beqa
K^{(1)}(z,0) = & & \rk{\ee^{\ii \frac{\eta}{2} H(z,0)} \frac{1}{2} \Lambda P^{(0)}(z,0)} \ee^{-\ii \frac{\eta}{2} H(z,0)} \nonumber\\
& & + \rk{\ee^{\ii \frac{\eta}{2} H(z,0)} P^{(0)}(z,0)} \frac{1}{2} \Lambda \ee^{-\ii \frac{\eta}{2} H(z,0)} \nonumber\\
= & & \frac{\ii \eta}{2} \half \sk{H(z,0) \Lambda P^{(0)}(z,0) - P^{(0)}(z,0) \Lambda H(z,0)} \nonumber\\
= & & \frac{\ii \eta}{2} H(z,0) \Lambda P^{(0)}(z,0).
\label{Kz0_1_FCS}
\eeqa
Equations (\ref{Kz0_0_FCS}) and (\ref{Kz0_1_FCS}) are \eqnref{Initial:FCS_WW}.

For the initial condition of $\hat{K}(t)$ based on $\Phi_{\text{MH}}(\eta)$ [\eqnref{Initial:MH}], the Weyl symbol of the operator $\hat{K}(0)$ is
\begin{widetext}
\beq
K(z,0) = \half \sk{P(z,0) + \rk{\bk{\ee^{\ii \eta \hat{H}(0)}}_w \exp \rk{\frac{\ii \hbar}{2} \Lambda} P(z,0)} \exp \rk{\frac{\ii \hbar}{2} \Lambda} \bk{\ee^{-\ii \eta \hat{H}(0)}}_w}.
\label{Kz0_MH}
\eeq
\end{widetext}
The zeroth order is 
\beqa
K^{(0)}(z,0) &=& \half \sk{P^{(0)}(z,0) + \ee^{\ii \eta H(z,0)} P^{(0)}(z,0) \ee^{-\ii \eta H(z,0)}} \nonumber\\
&=& P^{(0)}(z,0).
\label{Kz0_0_MH}
\eeqa
Using the conclusion of \eqnref{Kz0_1_FCS} , the first order is the same as that based on the FCS:
\beqa
K^{(1)}(z,0) &=& \half \sk{0 + \ii \eta H(z,0) \Lambda P^{(0)}(z,0)} \nonumber\\
&=& \frac{\ii \eta}{2} H(z,0) \Lambda P^{(0)}(z,0).
\label{Kz0_1_MH}
\eeqa
Equations (\ref{Kz0_0_MH}) and (\ref{Kz0_1_MH}) are also \eqnref{Initial:FCS_WW}.

We would like to emphasize that $K^{(2)}(z,0)$, $K^{(3)}(z,0)$, $\cdots$ based on $\Phi_{\text{FCS}}(\eta)$ and $\Phi_{\text{MH}}(\eta)$ are different [see \eqnref{FCSMH:Phi2}].

Then let us derive \eqnref{Mz0 expansion}. From the initial condition of $\hat{M}(t)$ based on $\Phi_{\text{FCS}}(\eta)$ [\eqnref{Initial:M_FCS}], we obtain the Weyl symbol of the operator $\hat{M}(0)$:
\begin{widetext}
\beq
M(z,0) = \sk{\bk{\ee^{-\ii \frac{\eta}{2} \hat{H}(0)}}_w \exp \rk{\frac{\ii \hbar}{2} \Lambda} P(z,0)} \exp \rk{\frac{\ii \hbar}{2} \Lambda} \bk{\ee^{-\ii \frac{\eta}{2} \hat{H}(0)}}_w.
\label{Mz0_FCS}
\eeq
\end{widetext}
The zeroth order is
\beq
M^{(0)}(z,0) = \ee^{-\ii \eta H(z,0)} P^{(0)}(z,0).
\label{Mz0_0_FCS}
\eeq
The first order originating from the two star products $\exp \rk{\frac{\ii \hbar}{2} \Lambda}$ is
\beqa
M^{(1)}(z,0) = & & \rk{\ee^{-\ii \frac{\eta}{2} H(z,0)} \frac{1}{2} \Lambda P^{(0)}(z,0)} \ee^{-\ii \frac{\eta}{2} H(z,0)} \nonumber\\
& & + \rk{\ee^{-\ii \frac{\eta}{2} H(z,0)} P^{(0)}(z,0)} \frac{1}{2} \Lambda \ee^{-\ii \frac{\eta}{2} H(z,0)} \nonumber\\
= & & -\frac{\ii \eta}{2} \ee^{-\ii \eta H(z,0)} \times \nonumber\\
& & \half \sk{H(z,0) \Lambda P^{(0)}(z,0) + P^{(0)}(z,0) \Lambda H(z,0)} \nonumber\\
= & & 0.
\label{Mz0_1_FCS}
\eeqa
Thus the first order vanishes. Similarly, the terms in $M(z,0)$ proportional to odd orders of $\hbar$ vanish.

For the initial condition of $\hat{M}(t)$ based on $\Phi_{\text{MH}}(\eta)$ [\eqnref{Initial:M_MH}], the Weyl symbol of the operator $\hat{M}(0)$ is
\begin{widetext}
\beqa
M(z,0) &=& \half \sk{\bk{\ee^{-\ii \eta \hat{H}(0)}}_w \exp \rk{\frac{\ii \hbar}{2} \Lambda} P(z,0) + P(z,0) \exp \rk{\frac{\ii \hbar}{2} \Lambda} \bk{\ee^{-\ii \eta \hat{H}(0)}}_w} \nonumber\\
&=& \bk{\ee^{-\ii \eta \hat{H}(0)}}_w \cos \rk{\frac{\hbar}{2} \Lambda} P(z,0).
\label{Mz0_MH}
\eeqa
\end{widetext}
The zeroth order is the same as \eqnref{Mz0_0_FCS}. The terms proportional to odd orders of $\hbar$ vanish because $\bk{\ee^{-\ii \eta \hat{H}(0)}}_w$, $\cos \rk{\frac{\hbar}{2} \Lambda}$, and $P(z,0)$ only contain terms proportional to even orders of $\hbar$. Thus for both work definitions based on the FCS and the MH, $M(z,0)$ can be expanded in the form of \eqnref{Mz0 expansion}.

We would like to emphasize that $M^{(2)}(z,0)$, $M^{(4)}(z,0)$, $\cdots$ based on $\Phi_{\text{FCS}}(\eta)$ and $\Phi_{\text{MH}}(\eta)$ are different [see \eqnref{FCSMH:Phi2}].

\section{\label{A:HO0}Calculation of $\Phi^{(0)}(\eta)$ and $\Phi^{(2)}(\eta)$ for the linearly dragged harmonic oscillator}

From \eqnref{Eqn:EOM_M_WW}, we know that both $M^{(0)}(z,t)$ and $M^{(2)}(z,t)$ satisfy the classical Liouville equation. That is, the trajectory in the phase space satisfies Newton's equation. For the Hamiltonian of the linearly dragged harmonic oscillator [\eqnref{Hamiltonian:HO}], we obtain
\beq
\left\lbrace \begin{aligned}
& x(t) = x(0) \cos \omega t + \frac{p(0)}{m \omega} \sin \omega t + u t - \frac{u}{\omega} \sin \omega t,\\
& p(t) = - x(0) m \omega \sin \omega t + p(0) \cos \omega t + m u (1 - \cos \omega t).
\end{aligned} \right.
\label{zt}
\eeq
This is a mapping from $\rk{x(0),p(0)}$ to $\rk{x(t),p(t)}$, and we denote it as 
\beq
z(t) = \psi_t \rk{z(0),0}.
\label{Map}
\eeq
The inverse mapping is
\beq
z(0) = \psi^{-1}_0 \rk{z(t),t}.
\label{Inverse map}
\eeq
More specifically,
\beq
\left\lbrace \begin{aligned}
& x(0) = x(t) \cos \omega t - \frac{p(t)}{m \omega} \sin \omega t - u \rk{t \cos \omega t - \frac{1}{\omega} \sin \omega t},\\
& p(0) = x(t) m \omega \sin \omega t + p(t) \cos \omega t - m u (\omega t \sin \omega t + \cos \omega t -1).
\end{aligned} \right.
\label{z0}
\eeq

Since in the isolated system, the Newton's trajectory starting from one phase space point is unique, the mapping [\eqnref{Inverse map}] is one-to-one. From Liouville's theorem, we obtain 
\beq
M^{(0)}(z,\tau) = M^{(0)} (\psi^{-1}_0(z,\tau),0).
\label{M0zt}
\eeq
By utilizing Eqs. (\ref{State:0}), (\ref{Mz0 expansion}) and (\ref{Eqn:CF_WW0}), we obtain
\begin{widetext}
\beqa
\Phi^{(0)}(\eta) &=& \int_{-\infty}^{+\infty} \dd z \ \ee^{\ii \eta H(z,\tau)} M^{(0)}(z,\tau) \nonumber\\
&=& \int_{-\infty}^{+\infty} \dd z \ \ee^{\ii \eta H(z,\tau)} \ee^{-\ii \eta H(\psi^{-1}_0(z,\tau),0)} P^{(0)}(\psi^{-1}_0(z,\tau),0) \nonumber\\
&=& \exp \bk{m u^2 \sk{\ii \eta \rk{\cos \omega \tau' - \cos \omega (\tau' + \tau)} - \frac{\eta^2}{\beta} \rk{1 - \cos \omega \tau}}}.
\label{M0zt_Calculation}
\eeqa
\end{widetext}
The derivation from the second line to the last line is not shown here since the integrals are all Gaussian. Equation (\ref{M0zt_Calculation}) is \eqnref{Sol:Phi0_HO}, which is identical to the classical characteristic function of work $\Phi_{\text{Classical}}(\eta)$ [\eqnref{Sol:PhiClassical_HO}].

Similarly, for the second order, we also have
\beq
M^{(2)}(z,\tau) = M^{(2)} (\psi^{-1}_0(z,\tau),0).
\label{M2zt}
\eeq
From \eqnref{Eqn:CF_WW}, we obtain that both $\bk{\ee^{\ii \eta \hat{H}(\tau)}}_w^{(2)}$ and $M^{(2)}(z,\tau)$ contribute to $\Phi^{(2)}(\eta)$:
\begin{widetext}
\beqa
\Phi^{(2)}(\eta) &=& \int_{-\infty}^{+\infty} \dd z \ \sk{\bk{\ee^{\ii \eta \hat{H}(\tau)}}_w^{(2)} M^{(0)}(z,\tau) + \ee^{\ii \eta H(z,\tau)} M^{(2)}(z,\tau)}\nonumber\\
&=& \int_{-\infty}^{+\infty} \dd z \ \sk{\ee^{\ii \eta H(z,\tau)} f(-\ii \eta,z,\tau) \ee^{-\ii \eta H(\psi^{-1}_0(z,\tau),0)} P^{(0)}(\psi^{-1}_0(z,\tau),0) + \ee^{\ii \eta H(z,\tau)} M^{(2)} (\psi^{-1}_0(z,\tau),0)},
\label{M2zt_Calculation}
\eeqa
\end{widetext}
where we have used Eqs. (\ref{Wigner expansion}) and (\ref{Def:f}). $M^{(2)}(z,0)$ can be calculated according to \eqnref{Mz0_FCS} (FCS) and \eqnref{Mz0_MH} (MH), respectively. The analytical form of \eqnref{M2zt_Calculation} is too complicated to show here, and we numerically calculate it as a function of $\eta$. The results are shown in \figref{}{Fig:1}.

\section{\label{A:Measurement}Derivation of \eqnref{WF measurement}}

For convenience, we set $m = \omega = 1$ in this section. As we know, after the first projective measurement, only the diagonal elements survive. The $n$-th diagonal element
\beqa
p_n &=& \Tr \sk{\hat{\rho}(0) \ket{n} \bra{n}} \nonumber\\
&=& \iint \dd x \dd p \ P(x,p,0) F_n(x,p),
\label{Diagonal}
\eeqa
where $F_n(x,p)$ is the Wigner function of the Fock state $\ket{n} \bra{n}$ \cite{barnett2002methods}:
\beq
F_n(x,p) = 2 (-1)^n \ee^{-2 \ab{\alpha}^2} L_n (4 \ab{\alpha}^2),
\label{Def:Fn}
\eeq
where 
\beq
\ab{\alpha}^2 = \frac{1}{2 \hbar} \rk{x^2 +p^2},
\label{Def:Alpha}
\eeq
and $L_n(x)$ is the Laguerre polynomials:
\beq
L_n(x) = \frac{\ee^x}{n!} \frac{\dd^n}{\dd x^n} \rk{\ee^{-x} x^n}.
\label{Def:Laguerre}
\eeq
After the projective measurement, the state becomes
\beq
\hat{\tilde{\rho}}(0) = \sum_{n=0}^{\infty} p_n \ket{n} \bra{n}.
\label{State:After measurement}
\eeq
Reformulating \eqnref{State:After measurement} in the Weyl-Wigner representation, the Wigner function of $\hat{\tilde{\rho}}(0)$ can be written as
\begin{widetext}
\beqa
\tilde{P}(x,p,0) &=& \frac{1}{2 \pi \hbar} \sum_{n=0}^{\infty} p_n F_n(x,p) \nonumber\\
&=& \frac{1}{2 \pi \hbar} \sum_{n=0}^{\infty} \sk{\iint \dd x' \dd p' \ P(x',p',0) F_n(x',p')} F_n(x,p) \nonumber\\
&=& \frac{2}{\pi \hbar} \iint \dd x' \dd p' \ P(x',p',0) \ee^{-\frac{1}{\hbar} \rk{x^2 + p^2 + x'^2 + p'^2}} \sum_{n=0}^{\infty} L_n \rk{\frac{2}{\hbar} \rk{x^2 + p^2}} L_n \rk{\frac{2}{\hbar} \rk{x'^2 + p'^2}}.
\label{WF:After measurement1}
\eeqa
\end{widetext}
Using the identity of the Laguerre polynomials \cite{barnett2002methods}
\beq
\sum_{n=0}^{\infty} L_n(x) L_n(y) = \ee^{\frac{x+y}{2}} \delta(x-y),
\label{Identity:Laguerre}
\eeq
and converting \eqnref{WF:After measurement1} into the polar coordinate, we obtain
\beqa
\tilde{P}(r,\theta,0) &=& \frac{2}{\pi \hbar} \int_0^{\infty} r' \dd r' \int_0^{2 \pi} \dd \theta' P(r',\theta',0) \delta \rk{\frac{2}{\hbar} \rk{r^2 - r'^2}} \nonumber\\
&=& \frac{1}{2 \pi} \int_0^{2 \pi} \dd \theta' P(r,\theta',0).
\label{WF:After measurement2}
\eeqa
This is \eqnref{WF measurement} (the r.h.s. of \eqnref{WF:After measurement2} is independent of $\theta$).

\bibliography{QCRefs}

\begin{thebibliography}{86}%
\makeatletter
\providecommand \@ifxundefined [1]{%
 \@ifx{#1\undefined}
}%
\providecommand \@ifnum [1]{%
 \ifnum #1\expandafter \@firstoftwo
 \else \expandafter \@secondoftwo
 \fi
}%
\providecommand \@ifx [1]{%
 \ifx #1\expandafter \@firstoftwo
 \else \expandafter \@secondoftwo
 \fi
}%
\providecommand \natexlab [1]{#1}%
\providecommand \enquote  [1]{``#1''}%
\providecommand \bibnamefont  [1]{#1}%
\providecommand \bibfnamefont [1]{#1}%
\providecommand \citenamefont [1]{#1}%
\providecommand \href@noop [0]{\@secondoftwo}%
\providecommand \href [0]{\begingroup \@sanitize@url \@href}%
\providecommand \@href[1]{\@@startlink{#1}\@@href}%
\providecommand \@@href[1]{\endgroup#1\@@endlink}%
\providecommand \@sanitize@url [0]{\catcode `\\12\catcode `\$12\catcode
  `\&12\catcode `\#12\catcode `\^12\catcode `\_12\catcode `\%12\relax}%
\providecommand \@@startlink[1]{}%
\providecommand \@@endlink[0]{}%
\providecommand \url  [0]{\begingroup\@sanitize@url \@url }%
\providecommand \@url [1]{\endgroup\@href {#1}{\urlprefix }}%
\providecommand \urlprefix  [0]{URL }%
\providecommand \Eprint [0]{\href }%
\providecommand \doibase [0]{http://dx.doi.org/}%
\providecommand \selectlanguage [0]{\@gobble}%
\providecommand \bibinfo  [0]{\@secondoftwo}%
\providecommand \bibfield  [0]{\@secondoftwo}%
\providecommand \translation [1]{[#1]}%
\providecommand \BibitemOpen [0]{}%
\providecommand \bibitemStop [0]{}%
\providecommand \bibitemNoStop [0]{.\EOS\space}%
\providecommand \EOS [0]{\spacefactor3000\relax}%
\providecommand \BibitemShut  [1]{\csname bibitem#1\endcsname}%
\let\auto@bib@innerbib\@empty
\bibitem [{\citenamefont {Jarzynski}(2011)}]{jarzynski2011equalities}%
  \BibitemOpen
  \bibfield  {author} {\bibinfo {author} {\bibfnamefont {C.}~\bibnamefont
  {Jarzynski}},\ }\href@noop {} {\bibfield  {journal} {\bibinfo  {journal}
  {Annu. Rev. Condens. Matter Phys.}\ }\textbf {\bibinfo {volume} {2}},\
  \bibinfo {pages} {329} (\bibinfo {year} {2011})}\BibitemShut {NoStop}%
\bibitem [{\citenamefont {Seifert}(2012)}]{seifert2012stochastic}%
  \BibitemOpen
  \bibfield  {author} {\bibinfo {author} {\bibfnamefont {U.}~\bibnamefont
  {Seifert}},\ }\href@noop {} {\bibfield  {journal} {\bibinfo  {journal}
  {Reports on Progress in Physics}\ }\textbf {\bibinfo {volume} {75}},\
  \bibinfo {pages} {126001} (\bibinfo {year} {2012})}\BibitemShut {NoStop}%
\bibitem [{\citenamefont {Sekimoto}(1998)}]{sekimoto1998langevin}%
  \BibitemOpen
  \bibfield  {author} {\bibinfo {author} {\bibfnamefont {K.}~\bibnamefont
  {Sekimoto}},\ }\href@noop {} {\bibfield  {journal} {\bibinfo  {journal}
  {Progress of Theoretical Physics Supplement}\ }\textbf {\bibinfo {volume}
  {130}},\ \bibinfo {pages} {17} (\bibinfo {year} {1998})}\BibitemShut
  {NoStop}%
\bibitem [{\citenamefont {Sekimoto}(2010)}]{sekimoto2010stochastic}%
  \BibitemOpen
  \bibfield  {author} {\bibinfo {author} {\bibfnamefont {K.}~\bibnamefont
  {Sekimoto}},\ }\href@noop {} {\emph {\bibinfo {title} {Stochastic
  Energetics}}},\ \bibinfo {series} {Lecture Notes in Physics}, Vol.\ \bibinfo
  {volume} {799}\ (\bibinfo  {publisher} {Springer-Verlag},\ \bibinfo {address}
  {Berlin},\ \bibinfo {year} {2010})\BibitemShut {NoStop}%
\bibitem [{\citenamefont {Li}\ \emph {et~al.}(2010)\citenamefont {Li},
  \citenamefont {Kheifets}, \citenamefont {Medellin},\ and\ \citenamefont
  {Raizen}}]{li2010measurement}%
  \BibitemOpen
  \bibfield  {author} {\bibinfo {author} {\bibfnamefont {T.}~\bibnamefont
  {Li}}, \bibinfo {author} {\bibfnamefont {S.}~\bibnamefont {Kheifets}},
  \bibinfo {author} {\bibfnamefont {D.}~\bibnamefont {Medellin}}, \ and\
  \bibinfo {author} {\bibfnamefont {M.~G.}\ \bibnamefont {Raizen}},\
  }\href@noop {} {\bibfield  {journal} {\bibinfo  {journal} {Science}\ }\textbf
  {\bibinfo {volume} {328}},\ \bibinfo {pages} {1673} (\bibinfo {year}
  {2010})}\BibitemShut {NoStop}%
\bibitem [{\citenamefont {Mart{\'\i}nez}\ \emph {et~al.}(2016)\citenamefont
  {Mart{\'\i}nez}, \citenamefont {Rold{\'a}n}, \citenamefont {Dinis},
  \citenamefont {Petrov}, \citenamefont {Parrondo},\ and\ \citenamefont
  {Rica}}]{martinez2016brownian}%
  \BibitemOpen
  \bibfield  {author} {\bibinfo {author} {\bibfnamefont {I.~A.}\ \bibnamefont
  {Mart{\'\i}nez}}, \bibinfo {author} {\bibfnamefont {{\'E}.}~\bibnamefont
  {Rold{\'a}n}}, \bibinfo {author} {\bibfnamefont {L.}~\bibnamefont {Dinis}},
  \bibinfo {author} {\bibfnamefont {D.}~\bibnamefont {Petrov}}, \bibinfo
  {author} {\bibfnamefont {J.~M.~R.}\ \bibnamefont {Parrondo}}, \ and\ \bibinfo
  {author} {\bibfnamefont {R.~A.}\ \bibnamefont {Rica}},\ }\href@noop {}
  {\bibfield  {journal} {\bibinfo  {journal} {Nature physics}\ }\textbf
  {\bibinfo {volume} {12}},\ \bibinfo {pages} {67} (\bibinfo {year}
  {2016})}\BibitemShut {NoStop}%
\bibitem [{\citenamefont {Bang}\ \emph {et~al.}(2018)\citenamefont {Bang},
  \citenamefont {Pan}, \citenamefont {Hoang}, \citenamefont {Ahn},
  \citenamefont {Jarzynski}, \citenamefont {Quan},\ and\ \citenamefont
  {Li}}]{bang2018experimental}%
  \BibitemOpen
  \bibfield  {author} {\bibinfo {author} {\bibfnamefont {J.}~\bibnamefont
  {Bang}}, \bibinfo {author} {\bibfnamefont {R.}~\bibnamefont {Pan}}, \bibinfo
  {author} {\bibfnamefont {T.~M.}\ \bibnamefont {Hoang}}, \bibinfo {author}
  {\bibfnamefont {J.}~\bibnamefont {Ahn}}, \bibinfo {author} {\bibfnamefont
  {C.}~\bibnamefont {Jarzynski}}, \bibinfo {author} {\bibfnamefont {H.~T.}\
  \bibnamefont {Quan}}, \ and\ \bibinfo {author} {\bibfnamefont
  {T.}~\bibnamefont {Li}},\ }\href@noop {} {\bibfield  {journal} {\bibinfo
  {journal} {New Journal of Physics}\ }\textbf {\bibinfo {volume} {20}},\
  \bibinfo {pages} {103032} (\bibinfo {year} {2018})}\BibitemShut {NoStop}%
\bibitem [{\citenamefont {Jarzynski}(1997)}]{jarzynski1997nonequilibrium}%
  \BibitemOpen
  \bibfield  {author} {\bibinfo {author} {\bibfnamefont {C.}~\bibnamefont
  {Jarzynski}},\ }\href@noop {} {\bibfield  {journal} {\bibinfo  {journal}
  {Physical Review Letters}\ }\textbf {\bibinfo {volume} {78}},\ \bibinfo
  {pages} {2690} (\bibinfo {year} {1997})}\BibitemShut {NoStop}%
\bibitem [{\citenamefont {Crooks}(1999)}]{crooks1999entropy}%
  \BibitemOpen
  \bibfield  {author} {\bibinfo {author} {\bibfnamefont {G.~E.}\ \bibnamefont
  {Crooks}},\ }\href@noop {} {\bibfield  {journal} {\bibinfo  {journal}
  {Physical Review E}\ }\textbf {\bibinfo {volume} {60}},\ \bibinfo {pages}
  {2721} (\bibinfo {year} {1999})}\BibitemShut {NoStop}%
\bibitem [{\citenamefont {Seifert}(2005)}]{seifert2005entropy}%
  \BibitemOpen
  \bibfield  {author} {\bibinfo {author} {\bibfnamefont {U.}~\bibnamefont
  {Seifert}},\ }\href@noop {} {\bibfield  {journal} {\bibinfo  {journal}
  {Physical Review Letters}\ }\textbf {\bibinfo {volume} {95}},\ \bibinfo
  {pages} {040602} (\bibinfo {year} {2005})}\BibitemShut {NoStop}%
\bibitem [{\citenamefont {Hummer}\ and\ \citenamefont
  {Szabo}(2001)}]{hummer2001free}%
  \BibitemOpen
  \bibfield  {author} {\bibinfo {author} {\bibfnamefont {G.}~\bibnamefont
  {Hummer}}\ and\ \bibinfo {author} {\bibfnamefont {A.}~\bibnamefont {Szabo}},\
  }\href@noop {} {\bibfield  {journal} {\bibinfo  {journal} {Proc. Natl. Acad.
  Sci.}\ }\textbf {\bibinfo {volume} {98}},\ \bibinfo {pages} {3658} (\bibinfo
  {year} {2001})}\BibitemShut {NoStop}%
\bibitem [{\citenamefont {Kawai}\ \emph {et~al.}(2007)\citenamefont {Kawai},
  \citenamefont {Parrondo},\ and\ \citenamefont {van~den
  Broeck}}]{kawai2007dissipation}%
  \BibitemOpen
  \bibfield  {author} {\bibinfo {author} {\bibfnamefont {R.}~\bibnamefont
  {Kawai}}, \bibinfo {author} {\bibfnamefont {J.~M.~R.}\ \bibnamefont
  {Parrondo}}, \ and\ \bibinfo {author} {\bibfnamefont {C.}~\bibnamefont
  {van~den Broeck}},\ }\href@noop {} {\bibfield  {journal} {\bibinfo  {journal}
  {Phys. Rev. Lett.}\ }\textbf {\bibinfo {volume} {98}},\ \bibinfo {pages}
  {080602} (\bibinfo {year} {2007})}\BibitemShut {NoStop}%
\bibitem [{\citenamefont {Esposito}\ and\ \citenamefont {van~den
  Broeck}(2010)}]{esposito2010detailed}%
  \BibitemOpen
  \bibfield  {author} {\bibinfo {author} {\bibfnamefont {M.}~\bibnamefont
  {Esposito}}\ and\ \bibinfo {author} {\bibfnamefont {C.}~\bibnamefont {van~den
  Broeck}},\ }\href@noop {} {\bibfield  {journal} {\bibinfo  {journal} {Phys.
  Rev. Lett.}\ }\textbf {\bibinfo {volume} {104}},\ \bibinfo {pages} {090601}
  (\bibinfo {year} {2010})}\BibitemShut {NoStop}%
\bibitem [{\citenamefont {Gong}\ and\ \citenamefont
  {Quan}(2015)}]{gong2015jarzynski}%
  \BibitemOpen
  \bibfield  {author} {\bibinfo {author} {\bibfnamefont {Z.}~\bibnamefont
  {Gong}}\ and\ \bibinfo {author} {\bibfnamefont {H.~T.}\ \bibnamefont
  {Quan}},\ }\href@noop {} {\bibfield  {journal} {\bibinfo  {journal} {Physical
  Review E}\ }\textbf {\bibinfo {volume} {92}},\ \bibinfo {pages} {012131}
  (\bibinfo {year} {2015})}\BibitemShut {NoStop}%
\bibitem [{\citenamefont {Liphardt}\ \emph {et~al.}(2002)\citenamefont
  {Liphardt}, \citenamefont {Dumont}, \citenamefont {Smith}, \citenamefont
  {Tinoco},\ and\ \citenamefont {Bustamante}}]{liphardt2002equilibrium}%
  \BibitemOpen
  \bibfield  {author} {\bibinfo {author} {\bibfnamefont {J.}~\bibnamefont
  {Liphardt}}, \bibinfo {author} {\bibfnamefont {S.}~\bibnamefont {Dumont}},
  \bibinfo {author} {\bibfnamefont {S.~B.}\ \bibnamefont {Smith}}, \bibinfo
  {author} {\bibfnamefont {I.}~\bibnamefont {Tinoco}}, \ and\ \bibinfo {author}
  {\bibfnamefont {C.}~\bibnamefont {Bustamante}},\ }\href@noop {} {\bibfield
  {journal} {\bibinfo  {journal} {Science}\ }\textbf {\bibinfo {volume}
  {296}},\ \bibinfo {pages} {1832} (\bibinfo {year} {2002})}\BibitemShut
  {NoStop}%
\bibitem [{\citenamefont {Collin}\ \emph {et~al.}(2005)\citenamefont {Collin},
  \citenamefont {Ritort}, \citenamefont {Jarzynski}, \citenamefont {Smith},
  \citenamefont {Tinoco},\ and\ \citenamefont
  {Bustamante}}]{collin2005verification}%
  \BibitemOpen
  \bibfield  {author} {\bibinfo {author} {\bibfnamefont {D.}~\bibnamefont
  {Collin}}, \bibinfo {author} {\bibfnamefont {F.}~\bibnamefont {Ritort}},
  \bibinfo {author} {\bibfnamefont {C.}~\bibnamefont {Jarzynski}}, \bibinfo
  {author} {\bibfnamefont {S.~B.}\ \bibnamefont {Smith}}, \bibinfo {author}
  {\bibfnamefont {I.}~\bibnamefont {Tinoco}}, \ and\ \bibinfo {author}
  {\bibfnamefont {C.}~\bibnamefont {Bustamante}},\ }\href@noop {} {\bibfield
  {journal} {\bibinfo  {journal} {Nature (London)}\ }\textbf {\bibinfo {volume}
  {437}},\ \bibinfo {pages} {231} (\bibinfo {year} {2005})}\BibitemShut
  {NoStop}%
\bibitem [{\citenamefont {Douarche}\ \emph {et~al.}(2006)\citenamefont
  {Douarche}, \citenamefont {Joubaud}, \citenamefont {Garnier}, \citenamefont
  {Petrosyan},\ and\ \citenamefont {Ciliberto}}]{douarche2006work}%
  \BibitemOpen
  \bibfield  {author} {\bibinfo {author} {\bibfnamefont {F.}~\bibnamefont
  {Douarche}}, \bibinfo {author} {\bibfnamefont {S.}~\bibnamefont {Joubaud}},
  \bibinfo {author} {\bibfnamefont {N.~B.}\ \bibnamefont {Garnier}}, \bibinfo
  {author} {\bibfnamefont {A.}~\bibnamefont {Petrosyan}}, \ and\ \bibinfo
  {author} {\bibfnamefont {S.}~\bibnamefont {Ciliberto}},\ }\href@noop {}
  {\bibfield  {journal} {\bibinfo  {journal} {Physical Review Letters}\
  }\textbf {\bibinfo {volume} {97}},\ \bibinfo {pages} {140603} (\bibinfo
  {year} {2006})}\BibitemShut {NoStop}%
\bibitem [{\citenamefont {Junier}\ \emph {et~al.}(2009)\citenamefont {Junier},
  \citenamefont {Mossa}, \citenamefont {Manosas},\ and\ \citenamefont
  {Ritort}}]{junier2009recovery}%
  \BibitemOpen
  \bibfield  {author} {\bibinfo {author} {\bibfnamefont {I.}~\bibnamefont
  {Junier}}, \bibinfo {author} {\bibfnamefont {A.}~\bibnamefont {Mossa}},
  \bibinfo {author} {\bibfnamefont {M.}~\bibnamefont {Manosas}}, \ and\
  \bibinfo {author} {\bibfnamefont {F.}~\bibnamefont {Ritort}},\ }\href@noop {}
  {\bibfield  {journal} {\bibinfo  {journal} {Physical Review Letters}\
  }\textbf {\bibinfo {volume} {102}},\ \bibinfo {pages} {070602} (\bibinfo
  {year} {2009})}\BibitemShut {NoStop}%
\bibitem [{\citenamefont {Toyabe}\ \emph {et~al.}(2010)\citenamefont {Toyabe},
  \citenamefont {Sagawa}, \citenamefont {Ueda}, \citenamefont {Muneyuki},\ and\
  \citenamefont {Sano}}]{toyabe2010experimental}%
  \BibitemOpen
  \bibfield  {author} {\bibinfo {author} {\bibfnamefont {S.}~\bibnamefont
  {Toyabe}}, \bibinfo {author} {\bibfnamefont {T.}~\bibnamefont {Sagawa}},
  \bibinfo {author} {\bibfnamefont {M.}~\bibnamefont {Ueda}}, \bibinfo {author}
  {\bibfnamefont {E.}~\bibnamefont {Muneyuki}}, \ and\ \bibinfo {author}
  {\bibfnamefont {M.}~\bibnamefont {Sano}},\ }\href@noop {} {\bibfield
  {journal} {\bibinfo  {journal} {Nature physics}\ }\textbf {\bibinfo {volume}
  {6}},\ \bibinfo {pages} {988} (\bibinfo {year} {2010})}\BibitemShut {NoStop}%
\bibitem [{\citenamefont {Pekola}(2015)}]{pekola2015review}%
  \BibitemOpen
  \bibfield  {author} {\bibinfo {author} {\bibfnamefont {J.~P.}\ \bibnamefont
  {Pekola}},\ }\href@noop {} {\bibfield  {journal} {\bibinfo  {journal} {Nature
  Phys.}\ }\textbf {\bibinfo {volume} {11}},\ \bibinfo {pages} {118} (\bibinfo
  {year} {2015})}\BibitemShut {NoStop}%
\bibitem [{\citenamefont {Hoang}\ \emph {et~al.}(2018)\citenamefont {Hoang},
  \citenamefont {Pan}, \citenamefont {Ahn}, \citenamefont {Bang}, \citenamefont
  {Quan},\ and\ \citenamefont {Li}}]{hoang2018experimental}%
  \BibitemOpen
  \bibfield  {author} {\bibinfo {author} {\bibfnamefont {T.~M.}\ \bibnamefont
  {Hoang}}, \bibinfo {author} {\bibfnamefont {R.}~\bibnamefont {Pan}}, \bibinfo
  {author} {\bibfnamefont {J.}~\bibnamefont {Ahn}}, \bibinfo {author}
  {\bibfnamefont {J.}~\bibnamefont {Bang}}, \bibinfo {author} {\bibfnamefont
  {H.~T.}\ \bibnamefont {Quan}}, \ and\ \bibinfo {author} {\bibfnamefont
  {T.}~\bibnamefont {Li}},\ }\href@noop {} {\bibfield  {journal} {\bibinfo
  {journal} {Physical Review Letters}\ }\textbf {\bibinfo {volume} {120}},\
  \bibinfo {pages} {080602} (\bibinfo {year} {2018})}\BibitemShut {NoStop}%
\bibitem [{\citenamefont {Talkner}\ and\ \citenamefont
  {H{\"a}nggi}(2016)}]{talkner2016aspects}%
  \BibitemOpen
  \bibfield  {author} {\bibinfo {author} {\bibfnamefont {P.}~\bibnamefont
  {Talkner}}\ and\ \bibinfo {author} {\bibfnamefont {P.}~\bibnamefont
  {H{\"a}nggi}},\ }\href@noop {} {\bibfield  {journal} {\bibinfo  {journal}
  {Physical Review E}\ }\textbf {\bibinfo {volume} {93}},\ \bibinfo {pages}
  {022131} (\bibinfo {year} {2016})}\BibitemShut {NoStop}%
\bibitem [{\citenamefont {Funo}\ and\ \citenamefont
  {Quan}(2018{\natexlab{a}})}]{funo2018path}%
  \BibitemOpen
  \bibfield  {author} {\bibinfo {author} {\bibfnamefont {K.}~\bibnamefont
  {Funo}}\ and\ \bibinfo {author} {\bibfnamefont {H.~T.}\ \bibnamefont
  {Quan}},\ }\href@noop {} {\bibfield  {journal} {\bibinfo  {journal} {Physical
  Review Letters}\ }\textbf {\bibinfo {volume} {121}},\ \bibinfo {pages}
  {040602} (\bibinfo {year} {2018}{\natexlab{a}})}\BibitemShut {NoStop}%
\bibitem [{\citenamefont {Funo}\ and\ \citenamefont
  {Quan}(2018{\natexlab{b}})}]{funo2018heat}%
  \BibitemOpen
  \bibfield  {author} {\bibinfo {author} {\bibfnamefont {K.}~\bibnamefont
  {Funo}}\ and\ \bibinfo {author} {\bibfnamefont {H.~T.}\ \bibnamefont
  {Quan}},\ }\href {\doibase 10.1103/PhysRevE.98.012113} {\bibfield  {journal}
  {\bibinfo  {journal} {Phys. Rev. E}\ }\textbf {\bibinfo {volume} {98}},\
  \bibinfo {pages} {012113} (\bibinfo {year} {2018}{\natexlab{b}})}\BibitemShut
  {NoStop}%
\bibitem [{\citenamefont {Kurchan}(2000)}]{kurchan2000quantum}%
  \BibitemOpen
  \bibfield  {author} {\bibinfo {author} {\bibfnamefont {J.}~\bibnamefont
  {Kurchan}},\ }\href@noop {} {\bibfield  {journal} {\bibinfo  {journal}
  {arXiv:cond-mat/0007360}\ } (\bibinfo {year} {2000})}\BibitemShut {NoStop}%
\bibitem [{\citenamefont {Tasaki}(2000)}]{tasaki2000jarzynski}%
  \BibitemOpen
  \bibfield  {author} {\bibinfo {author} {\bibfnamefont {H.}~\bibnamefont
  {Tasaki}},\ }\href@noop {} {\bibfield  {journal} {\bibinfo  {journal}
  {arXiv:cond-mat/0009244}\ } (\bibinfo {year} {2000})}\BibitemShut {NoStop}%
\bibitem [{\citenamefont {Talkner}\ \emph {et~al.}(2007)\citenamefont
  {Talkner}, \citenamefont {Lutz},\ and\ \citenamefont
  {H{\"a}nggi}}]{talkner2007fluctuation}%
  \BibitemOpen
  \bibfield  {author} {\bibinfo {author} {\bibfnamefont {P.}~\bibnamefont
  {Talkner}}, \bibinfo {author} {\bibfnamefont {E.}~\bibnamefont {Lutz}}, \
  and\ \bibinfo {author} {\bibfnamefont {P.}~\bibnamefont {H{\"a}nggi}},\
  }\href@noop {} {\bibfield  {journal} {\bibinfo  {journal} {Physical Review
  E}\ }\textbf {\bibinfo {volume} {75}},\ \bibinfo {pages} {050102} (\bibinfo
  {year} {2007})}\BibitemShut {NoStop}%
\bibitem [{\citenamefont {Brandner}\ and\ \citenamefont
  {Seifert}(2016)}]{PhysRevE.93.062134}%
  \BibitemOpen
  \bibfield  {author} {\bibinfo {author} {\bibfnamefont {K.}~\bibnamefont
  {Brandner}}\ and\ \bibinfo {author} {\bibfnamefont {U.}~\bibnamefont
  {Seifert}},\ }\href {\doibase 10.1103/PhysRevE.93.062134} {\bibfield
  {journal} {\bibinfo  {journal} {Phys. Rev. E}\ }\textbf {\bibinfo {volume}
  {93}},\ \bibinfo {pages} {062134} (\bibinfo {year} {2016})}\BibitemShut
  {NoStop}%
\bibitem [{\citenamefont {Suomela}\ \emph {et~al.}(2016)\citenamefont
  {Suomela}, \citenamefont {Kutvonen},\ and\ \citenamefont
  {Ala-Nissila}}]{PhysRevE.93.062106}%
  \BibitemOpen
  \bibfield  {author} {\bibinfo {author} {\bibfnamefont {S.}~\bibnamefont
  {Suomela}}, \bibinfo {author} {\bibfnamefont {A.}~\bibnamefont {Kutvonen}}, \
  and\ \bibinfo {author} {\bibfnamefont {T.}~\bibnamefont {Ala-Nissila}},\
  }\href {\doibase 10.1103/PhysRevE.93.062106} {\bibfield  {journal} {\bibinfo
  {journal} {Phys. Rev. E}\ }\textbf {\bibinfo {volume} {93}},\ \bibinfo
  {pages} {062106} (\bibinfo {year} {2016})}\BibitemShut {NoStop}%
\bibitem [{\citenamefont {Liu}(2016)}]{PhysRevE.93.012127}%
  \BibitemOpen
  \bibfield  {author} {\bibinfo {author} {\bibfnamefont {F.}~\bibnamefont
  {Liu}},\ }\href {\doibase 10.1103/PhysRevE.93.012127} {\bibfield  {journal}
  {\bibinfo  {journal} {Phys. Rev. E}\ }\textbf {\bibinfo {volume} {93}},\
  \bibinfo {pages} {012127} (\bibinfo {year} {2016})}\BibitemShut {NoStop}%
\bibitem [{\citenamefont {Wang}(2018)}]{PhysRevE.97.012128}%
  \BibitemOpen
  \bibfield  {author} {\bibinfo {author} {\bibfnamefont {W.-g.}\ \bibnamefont
  {Wang}},\ }\href {\doibase 10.1103/PhysRevE.97.012128} {\bibfield  {journal}
  {\bibinfo  {journal} {Phys. Rev. E}\ }\textbf {\bibinfo {volume} {97}},\
  \bibinfo {pages} {012128} (\bibinfo {year} {2018})}\BibitemShut {NoStop}%
\bibitem [{\citenamefont {Liu}\ and\ \citenamefont
  {Xi}(2016)}]{PhysRevE.94.062133}%
  \BibitemOpen
  \bibfield  {author} {\bibinfo {author} {\bibfnamefont {F.}~\bibnamefont
  {Liu}}\ and\ \bibinfo {author} {\bibfnamefont {J.}~\bibnamefont {Xi}},\
  }\href {\doibase 10.1103/PhysRevE.94.062133} {\bibfield  {journal} {\bibinfo
  {journal} {Phys. Rev. E}\ }\textbf {\bibinfo {volume} {94}},\ \bibinfo
  {pages} {062133} (\bibinfo {year} {2016})}\BibitemShut {NoStop}%
\bibitem [{\citenamefont {Rogers}(2017)}]{PhysRevE.95.012149}%
  \BibitemOpen
  \bibfield  {author} {\bibinfo {author} {\bibfnamefont {D.~M.}\ \bibnamefont
  {Rogers}},\ }\href {\doibase 10.1103/PhysRevE.95.012149} {\bibfield
  {journal} {\bibinfo  {journal} {Phys. Rev. E}\ }\textbf {\bibinfo {volume}
  {95}},\ \bibinfo {pages} {012149} (\bibinfo {year} {2017})}\BibitemShut
  {NoStop}%
\bibitem [{\citenamefont {Aurell}(2018)}]{PhysRevE.97.062117}%
  \BibitemOpen
  \bibfield  {author} {\bibinfo {author} {\bibfnamefont {E.}~\bibnamefont
  {Aurell}},\ }\href {\doibase 10.1103/PhysRevE.97.062117} {\bibfield
  {journal} {\bibinfo  {journal} {Phys. Rev. E}\ }\textbf {\bibinfo {volume}
  {97}},\ \bibinfo {pages} {062117} (\bibinfo {year} {2018})}\BibitemShut
  {NoStop}%
\bibitem [{\citenamefont {Sampaio}\ \emph {et~al.}(2016)\citenamefont
  {Sampaio}, \citenamefont {Suomela},\ and\ \citenamefont
  {Ala-Nissila}}]{PhysRevE.94.062122}%
  \BibitemOpen
  \bibfield  {author} {\bibinfo {author} {\bibfnamefont {R.}~\bibnamefont
  {Sampaio}}, \bibinfo {author} {\bibfnamefont {S.}~\bibnamefont {Suomela}}, \
  and\ \bibinfo {author} {\bibfnamefont {T.}~\bibnamefont {Ala-Nissila}},\
  }\href {\doibase 10.1103/PhysRevE.94.062122} {\bibfield  {journal} {\bibinfo
  {journal} {Phys. Rev. E}\ }\textbf {\bibinfo {volume} {94}},\ \bibinfo
  {pages} {062122} (\bibinfo {year} {2016})}\BibitemShut {NoStop}%
\bibitem [{\citenamefont {Jaramillo}\ \emph {et~al.}(2017)\citenamefont
  {Jaramillo}, \citenamefont {Deng},\ and\ \citenamefont
  {Gong}}]{PhysRevE.96.042119}%
  \BibitemOpen
  \bibfield  {author} {\bibinfo {author} {\bibfnamefont {J.~D.}\ \bibnamefont
  {Jaramillo}}, \bibinfo {author} {\bibfnamefont {J.}~\bibnamefont {Deng}}, \
  and\ \bibinfo {author} {\bibfnamefont {J.}~\bibnamefont {Gong}},\ }\href
  {\doibase 10.1103/PhysRevE.96.042119} {\bibfield  {journal} {\bibinfo
  {journal} {Phys. Rev. E}\ }\textbf {\bibinfo {volume} {96}},\ \bibinfo
  {pages} {042119} (\bibinfo {year} {2017})}\BibitemShut {NoStop}%
\bibitem [{\citenamefont {Gong}\ \emph {et~al.}(2016)\citenamefont {Gong},
  \citenamefont {Ashida},\ and\ \citenamefont {Ueda}}]{PhysRevA.94.012107}%
  \BibitemOpen
  \bibfield  {author} {\bibinfo {author} {\bibfnamefont {Z.}~\bibnamefont
  {Gong}}, \bibinfo {author} {\bibfnamefont {Y.}~\bibnamefont {Ashida}}, \ and\
  \bibinfo {author} {\bibfnamefont {M.}~\bibnamefont {Ueda}},\ }\href {\doibase
  10.1103/PhysRevA.94.012107} {\bibfield  {journal} {\bibinfo  {journal} {Phys.
  Rev. A}\ }\textbf {\bibinfo {volume} {94}},\ \bibinfo {pages} {012107}
  (\bibinfo {year} {2016})}\BibitemShut {NoStop}%
\bibitem [{\citenamefont {Talarico}\ \emph {et~al.}(2016)\citenamefont
  {Talarico}, \citenamefont {Monteiro}, \citenamefont {Mattei}, \citenamefont
  {Duzzioni}, \citenamefont {Souto~Ribeiro},\ and\ \citenamefont
  {C\'eleri}}]{PhysRevA.94.042305}%
  \BibitemOpen
  \bibfield  {author} {\bibinfo {author} {\bibfnamefont {M.~A.~A.}\
  \bibnamefont {Talarico}}, \bibinfo {author} {\bibfnamefont {P.~B.}\
  \bibnamefont {Monteiro}}, \bibinfo {author} {\bibfnamefont {E.~C.}\
  \bibnamefont {Mattei}}, \bibinfo {author} {\bibfnamefont {E.~I.}\
  \bibnamefont {Duzzioni}}, \bibinfo {author} {\bibfnamefont {P.~H.}\
  \bibnamefont {Souto~Ribeiro}}, \ and\ \bibinfo {author} {\bibfnamefont
  {L.~C.}\ \bibnamefont {C\'eleri}},\ }\href {\doibase
  10.1103/PhysRevA.94.042305} {\bibfield  {journal} {\bibinfo  {journal} {Phys.
  Rev. A}\ }\textbf {\bibinfo {volume} {94}},\ \bibinfo {pages} {042305}
  (\bibinfo {year} {2016})}\BibitemShut {NoStop}%
\bibitem [{\citenamefont {Bartolotta}\ and\ \citenamefont
  {Deffner}(2018)}]{PhysRevX.8.011033}%
  \BibitemOpen
  \bibfield  {author} {\bibinfo {author} {\bibfnamefont {A.}~\bibnamefont
  {Bartolotta}}\ and\ \bibinfo {author} {\bibfnamefont {S.}~\bibnamefont
  {Deffner}},\ }\href {\doibase 10.1103/PhysRevX.8.011033} {\bibfield
  {journal} {\bibinfo  {journal} {Phys. Rev. X}\ }\textbf {\bibinfo {volume}
  {8}},\ \bibinfo {pages} {011033} (\bibinfo {year} {2018})}\BibitemShut
  {NoStop}%
\bibitem [{\citenamefont {Silveri}\ \emph {et~al.}(2017)\citenamefont
  {Silveri}, \citenamefont {Tuorila}, \citenamefont {Thuneberg},\ and\
  \citenamefont {Paraoanu}}]{silveri2017quantum}%
  \BibitemOpen
  \bibfield  {author} {\bibinfo {author} {\bibfnamefont {M.}~\bibnamefont
  {Silveri}}, \bibinfo {author} {\bibfnamefont {J.}~\bibnamefont {Tuorila}},
  \bibinfo {author} {\bibfnamefont {E.}~\bibnamefont {Thuneberg}}, \ and\
  \bibinfo {author} {\bibfnamefont {G.}~\bibnamefont {Paraoanu}},\ }\href@noop
  {} {\bibfield  {journal} {\bibinfo  {journal} {Reports on Progress in
  Physics}\ }\textbf {\bibinfo {volume} {80}},\ \bibinfo {pages} {056002}
  (\bibinfo {year} {2017})}\BibitemShut {NoStop}%
\bibitem [{\citenamefont {Frenzel}\ \emph {et~al.}(2016)\citenamefont
  {Frenzel}, \citenamefont {Jennings},\ and\ \citenamefont
  {Rudolph}}]{frenzel2016quasi}%
  \BibitemOpen
  \bibfield  {author} {\bibinfo {author} {\bibfnamefont {M.~F.}\ \bibnamefont
  {Frenzel}}, \bibinfo {author} {\bibfnamefont {D.}~\bibnamefont {Jennings}}, \
  and\ \bibinfo {author} {\bibfnamefont {T.}~\bibnamefont {Rudolph}},\
  }\href@noop {} {\bibfield  {journal} {\bibinfo  {journal} {New Journal of
  Physics}\ }\textbf {\bibinfo {volume} {18}},\ \bibinfo {pages} {023037}
  (\bibinfo {year} {2016})}\BibitemShut {NoStop}%
\bibitem [{\citenamefont {Guarnieri}\ \emph {et~al.}(2018)\citenamefont
  {Guarnieri}, \citenamefont {Ng}, \citenamefont {Modi}, \citenamefont
  {Eisert}, \citenamefont {Paternostro},\ and\ \citenamefont
  {Goold}}]{guarnieri2018quantum}%
  \BibitemOpen
  \bibfield  {author} {\bibinfo {author} {\bibfnamefont {G.}~\bibnamefont
  {Guarnieri}}, \bibinfo {author} {\bibfnamefont {N.~H.~Y.}\ \bibnamefont
  {Ng}}, \bibinfo {author} {\bibfnamefont {K.}~\bibnamefont {Modi}}, \bibinfo
  {author} {\bibfnamefont {J.}~\bibnamefont {Eisert}}, \bibinfo {author}
  {\bibfnamefont {M.}~\bibnamefont {Paternostro}}, \ and\ \bibinfo {author}
  {\bibfnamefont {J.}~\bibnamefont {Goold}},\ }\href@noop {} {\bibfield
  {journal} {\bibinfo  {journal} {arXiv:1804.09962}\ } (\bibinfo {year}
  {2018})}\BibitemShut {NoStop}%
\bibitem [{\citenamefont {Kwon}\ and\ \citenamefont
  {Kim}(2018)}]{kwon2018fluctuation}%
  \BibitemOpen
  \bibfield  {author} {\bibinfo {author} {\bibfnamefont {H.}~\bibnamefont
  {Kwon}}\ and\ \bibinfo {author} {\bibfnamefont {M.}~\bibnamefont {Kim}},\
  }\href@noop {} {\bibfield  {journal} {\bibinfo  {journal} {arXiv:1810.03150}\
  } (\bibinfo {year} {2018})}\BibitemShut {NoStop}%
\bibitem [{\citenamefont {Peng}\ and\ \citenamefont
  {Fan}(2017)}]{peng2017perturbative}%
  \BibitemOpen
  \bibfield  {author} {\bibinfo {author} {\bibfnamefont {Y.}~\bibnamefont
  {Peng}}\ and\ \bibinfo {author} {\bibfnamefont {H.}~\bibnamefont {Fan}},\
  }\href@noop {} {\bibfield  {journal} {\bibinfo  {journal} {arXiv:1708.08214}\
  } (\bibinfo {year} {2017})}\BibitemShut {NoStop}%
\bibitem [{\citenamefont {Strasberg}(2018)}]{strasberg2018operational}%
  \BibitemOpen
  \bibfield  {author} {\bibinfo {author} {\bibfnamefont {P.}~\bibnamefont
  {Strasberg}},\ }\href@noop {} {\bibfield  {journal} {\bibinfo  {journal}
  {arXiv:1810.00698}\ } (\bibinfo {year} {2018})}\BibitemShut {NoStop}%
\bibitem [{\citenamefont {Goold}\ \emph {et~al.}(2016)\citenamefont {Goold},
  \citenamefont {Huber}, \citenamefont {Riera}, \citenamefont {del Rio},\ and\
  \citenamefont {Skrzypczyk}}]{goold2016role}%
  \BibitemOpen
  \bibfield  {author} {\bibinfo {author} {\bibfnamefont {J.}~\bibnamefont
  {Goold}}, \bibinfo {author} {\bibfnamefont {M.}~\bibnamefont {Huber}},
  \bibinfo {author} {\bibfnamefont {A.}~\bibnamefont {Riera}}, \bibinfo
  {author} {\bibfnamefont {L.}~\bibnamefont {del Rio}}, \ and\ \bibinfo
  {author} {\bibfnamefont {P.}~\bibnamefont {Skrzypczyk}},\ }\href@noop {}
  {\bibfield  {journal} {\bibinfo  {journal} {Journal of Physics A:
  Mathematical and Theoretical}\ }\textbf {\bibinfo {volume} {49}},\ \bibinfo
  {pages} {143001} (\bibinfo {year} {2016})}\BibitemShut {NoStop}%
\bibitem [{\citenamefont {Huber}\ \emph {et~al.}(2008)\citenamefont {Huber},
  \citenamefont {Schmidt-Kaler}, \citenamefont {Deffner},\ and\ \citenamefont
  {Lutz}}]{huber2008employing}%
  \BibitemOpen
  \bibfield  {author} {\bibinfo {author} {\bibfnamefont {G.}~\bibnamefont
  {Huber}}, \bibinfo {author} {\bibfnamefont {F.}~\bibnamefont
  {Schmidt-Kaler}}, \bibinfo {author} {\bibfnamefont {S.}~\bibnamefont
  {Deffner}}, \ and\ \bibinfo {author} {\bibfnamefont {E.}~\bibnamefont
  {Lutz}},\ }\href {\doibase 10.1103/PhysRevLett.101.070403} {\bibfield
  {journal} {\bibinfo  {journal} {Phys. Rev. Lett.}\ }\textbf {\bibinfo
  {volume} {101}},\ \bibinfo {pages} {070403} (\bibinfo {year}
  {2008})}\BibitemShut {NoStop}%
\bibitem [{\citenamefont {Batalh\~ao}\ \emph {et~al.}(2014)\citenamefont
  {Batalh\~ao}, \citenamefont {Souza}, \citenamefont {Mazzola}, \citenamefont
  {Auccaise}, \citenamefont {Sarthour}, \citenamefont {Oliveira}, \citenamefont
  {Goold}, \citenamefont {De~Chiara}, \citenamefont {Paternostro},\ and\
  \citenamefont {Serra}}]{batalhao2014experimental}%
  \BibitemOpen
  \bibfield  {author} {\bibinfo {author} {\bibfnamefont {T.~B.}\ \bibnamefont
  {Batalh\~ao}}, \bibinfo {author} {\bibfnamefont {A.~M.}\ \bibnamefont
  {Souza}}, \bibinfo {author} {\bibfnamefont {L.}~\bibnamefont {Mazzola}},
  \bibinfo {author} {\bibfnamefont {R.}~\bibnamefont {Auccaise}}, \bibinfo
  {author} {\bibfnamefont {R.~S.}\ \bibnamefont {Sarthour}}, \bibinfo {author}
  {\bibfnamefont {I.~S.}\ \bibnamefont {Oliveira}}, \bibinfo {author}
  {\bibfnamefont {J.}~\bibnamefont {Goold}}, \bibinfo {author} {\bibfnamefont
  {G.}~\bibnamefont {De~Chiara}}, \bibinfo {author} {\bibfnamefont
  {M.}~\bibnamefont {Paternostro}}, \ and\ \bibinfo {author} {\bibfnamefont
  {R.~M.}\ \bibnamefont {Serra}},\ }\href {\doibase
  10.1103/PhysRevLett.113.140601} {\bibfield  {journal} {\bibinfo  {journal}
  {Phys. Rev. Lett.}\ }\textbf {\bibinfo {volume} {113}},\ \bibinfo {pages}
  {140601} (\bibinfo {year} {2014})}\BibitemShut {NoStop}%
\bibitem [{\citenamefont {An}\ \emph {et~al.}(2015)\citenamefont {An},
  \citenamefont {Zhang}, \citenamefont {Um}, \citenamefont {Lv}, \citenamefont
  {Lu}, \citenamefont {Zhang}, \citenamefont {Yin}, \citenamefont {Quan},\ and\
  \citenamefont {Kim}}]{an2015experimental}%
  \BibitemOpen
  \bibfield  {author} {\bibinfo {author} {\bibfnamefont {S.}~\bibnamefont
  {An}}, \bibinfo {author} {\bibfnamefont {J.-N.}\ \bibnamefont {Zhang}},
  \bibinfo {author} {\bibfnamefont {M.}~\bibnamefont {Um}}, \bibinfo {author}
  {\bibfnamefont {D.}~\bibnamefont {Lv}}, \bibinfo {author} {\bibfnamefont
  {Y.}~\bibnamefont {Lu}}, \bibinfo {author} {\bibfnamefont {J.}~\bibnamefont
  {Zhang}}, \bibinfo {author} {\bibfnamefont {Z.-Q.}\ \bibnamefont {Yin}},
  \bibinfo {author} {\bibfnamefont {H.~T.}\ \bibnamefont {Quan}}, \ and\
  \bibinfo {author} {\bibfnamefont {K.}~\bibnamefont {Kim}},\ }\href@noop {}
  {\bibfield  {journal} {\bibinfo  {journal} {Nature Physics}\ }\textbf
  {\bibinfo {volume} {11}},\ \bibinfo {pages} {193} (\bibinfo {year}
  {2015})}\BibitemShut {NoStop}%
\bibitem [{\citenamefont {Jarzynski}\ \emph {et~al.}(2015)\citenamefont
  {Jarzynski}, \citenamefont {Quan},\ and\ \citenamefont
  {Rahav}}]{jarzynski2015quantum}%
  \BibitemOpen
  \bibfield  {author} {\bibinfo {author} {\bibfnamefont {C.}~\bibnamefont
  {Jarzynski}}, \bibinfo {author} {\bibfnamefont {H.~T.}\ \bibnamefont {Quan}},
  \ and\ \bibinfo {author} {\bibfnamefont {S.}~\bibnamefont {Rahav}},\
  }\href@noop {} {\bibfield  {journal} {\bibinfo  {journal} {Physical Review
  X}\ }\textbf {\bibinfo {volume} {5}},\ \bibinfo {pages} {031038} (\bibinfo
  {year} {2015})}\BibitemShut {NoStop}%
\bibitem [{\citenamefont {Zhu}\ \emph {et~al.}(2016)\citenamefont {Zhu},
  \citenamefont {Gong}, \citenamefont {Wu},\ and\ \citenamefont
  {Quan}}]{zhu2016quantum}%
  \BibitemOpen
  \bibfield  {author} {\bibinfo {author} {\bibfnamefont {L.}~\bibnamefont
  {Zhu}}, \bibinfo {author} {\bibfnamefont {Z.}~\bibnamefont {Gong}}, \bibinfo
  {author} {\bibfnamefont {B.}~\bibnamefont {Wu}}, \ and\ \bibinfo {author}
  {\bibfnamefont {H.~T.}\ \bibnamefont {Quan}},\ }\href@noop {} {\bibfield
  {journal} {\bibinfo  {journal} {Physical Review E}\ }\textbf {\bibinfo
  {volume} {93}},\ \bibinfo {pages} {062108} (\bibinfo {year}
  {2016})}\BibitemShut {NoStop}%
\bibitem [{\citenamefont {Wang}\ and\ \citenamefont
  {Quan}(2017)}]{wang2017understanding}%
  \BibitemOpen
  \bibfield  {author} {\bibinfo {author} {\bibfnamefont {Q.}~\bibnamefont
  {Wang}}\ and\ \bibinfo {author} {\bibfnamefont {H.~T.}\ \bibnamefont
  {Quan}},\ }\href {\doibase 10.1103/PhysRevE.95.032113} {\bibfield  {journal}
  {\bibinfo  {journal} {Phys. Rev. E}\ }\textbf {\bibinfo {volume} {95}},\
  \bibinfo {pages} {032113} (\bibinfo {year} {2017})}\BibitemShut {NoStop}%
\bibitem [{\citenamefont {Fei}\ \emph {et~al.}(2018)\citenamefont {Fei},
  \citenamefont {Quan},\ and\ \citenamefont {Liu}}]{fei2018quantum}%
  \BibitemOpen
  \bibfield  {author} {\bibinfo {author} {\bibfnamefont {Z.}~\bibnamefont
  {Fei}}, \bibinfo {author} {\bibfnamefont {H.~T.}\ \bibnamefont {Quan}}, \
  and\ \bibinfo {author} {\bibfnamefont {F.}~\bibnamefont {Liu}},\ }\href@noop
  {} {\bibfield  {journal} {\bibinfo  {journal} {Physical Review E}\ }\textbf
  {\bibinfo {volume} {98}},\ \bibinfo {pages} {012132} (\bibinfo {year}
  {2018})}\BibitemShut {NoStop}%
\bibitem [{\citenamefont {Garc\'{\i}a-Mata}\ \emph {et~al.}(2017)\citenamefont
  {Garc\'{\i}a-Mata}, \citenamefont {Roncaglia},\ and\ \citenamefont
  {Wisniacki}}]{garcia2017quantum}%
  \BibitemOpen
  \bibfield  {author} {\bibinfo {author} {\bibfnamefont {I.}~\bibnamefont
  {Garc\'{\i}a-Mata}}, \bibinfo {author} {\bibfnamefont {A.~J.}\ \bibnamefont
  {Roncaglia}}, \ and\ \bibinfo {author} {\bibfnamefont {D.~A.}\ \bibnamefont
  {Wisniacki}},\ }\href {\doibase 10.1103/PhysRevE.95.050102} {\bibfield
  {journal} {\bibinfo  {journal} {Phys. Rev. E}\ }\textbf {\bibinfo {volume}
  {95}},\ \bibinfo {pages} {050102} (\bibinfo {year} {2017})}\BibitemShut
  {NoStop}%
\bibitem [{\citenamefont {Arrais}\ \emph {et~al.}(2018)\citenamefont {Arrais},
  \citenamefont {Wisniacki}, \citenamefont {C\'eleri}, \citenamefont
  {de~Almeida}, \citenamefont {Roncaglia},\ and\ \citenamefont
  {Toscano}}]{arrais2018quantum}%
  \BibitemOpen
  \bibfield  {author} {\bibinfo {author} {\bibfnamefont {E.~G.}\ \bibnamefont
  {Arrais}}, \bibinfo {author} {\bibfnamefont {D.~A.}\ \bibnamefont
  {Wisniacki}}, \bibinfo {author} {\bibfnamefont {L.~C.}\ \bibnamefont
  {C\'eleri}}, \bibinfo {author} {\bibfnamefont {N.~G.}\ \bibnamefont
  {de~Almeida}}, \bibinfo {author} {\bibfnamefont {A.~J.}\ \bibnamefont
  {Roncaglia}}, \ and\ \bibinfo {author} {\bibfnamefont {F.}~\bibnamefont
  {Toscano}},\ }\href {\doibase 10.1103/PhysRevE.98.012106} {\bibfield
  {journal} {\bibinfo  {journal} {Phys. Rev. E}\ }\textbf {\bibinfo {volume}
  {98}},\ \bibinfo {pages} {012106} (\bibinfo {year} {2018})}\BibitemShut
  {NoStop}%
\bibitem [{\citenamefont {Garc{\'\i}a-Mata}\ \emph {et~al.}(2018)\citenamefont
  {Garc{\'\i}a-Mata}, \citenamefont {Roncaglia},\ and\ \citenamefont
  {Wisniacki}}]{garcia2018semiclassical}%
  \BibitemOpen
  \bibfield  {author} {\bibinfo {author} {\bibfnamefont {I.}~\bibnamefont
  {Garc{\'\i}a-Mata}}, \bibinfo {author} {\bibfnamefont {A.~J.}\ \bibnamefont
  {Roncaglia}}, \ and\ \bibinfo {author} {\bibfnamefont {D.~A.}\ \bibnamefont
  {Wisniacki}},\ }\href@noop {} {\bibfield  {journal} {\bibinfo  {journal} {EPL
  (Europhysics Letters)}\ }\textbf {\bibinfo {volume} {120}},\ \bibinfo {pages}
  {30002} (\bibinfo {year} {2018})}\BibitemShut {NoStop}%
\bibitem [{\citenamefont {Perarnau-Llobet}\ \emph {et~al.}(2017)\citenamefont
  {Perarnau-Llobet}, \citenamefont {B{\"a}umer}, \citenamefont {Hovhannisyan},
  \citenamefont {Huber},\ and\ \citenamefont {Acin}}]{perarnau2017no}%
  \BibitemOpen
  \bibfield  {author} {\bibinfo {author} {\bibfnamefont {M.}~\bibnamefont
  {Perarnau-Llobet}}, \bibinfo {author} {\bibfnamefont {E.}~\bibnamefont
  {B{\"a}umer}}, \bibinfo {author} {\bibfnamefont {K.~V.}\ \bibnamefont
  {Hovhannisyan}}, \bibinfo {author} {\bibfnamefont {M.}~\bibnamefont {Huber}},
  \ and\ \bibinfo {author} {\bibfnamefont {A.}~\bibnamefont {Acin}},\
  }\href@noop {} {\bibfield  {journal} {\bibinfo  {journal} {Physical Review
  Letters}\ }\textbf {\bibinfo {volume} {118}},\ \bibinfo {pages} {070601}
  (\bibinfo {year} {2017})}\BibitemShut {NoStop}%
\bibitem [{\citenamefont {Lostaglio}(2018)}]{lostaglio2018quantum}%
  \BibitemOpen
  \bibfield  {author} {\bibinfo {author} {\bibfnamefont {M.}~\bibnamefont
  {Lostaglio}},\ }\href@noop {} {\bibfield  {journal} {\bibinfo  {journal}
  {Physical Review Letters}\ }\textbf {\bibinfo {volume} {120}},\ \bibinfo
  {pages} {040602} (\bibinfo {year} {2018})}\BibitemShut {NoStop}%
\bibitem [{\citenamefont {B{\"a}umer}\ \emph {et~al.}(2018)\citenamefont
  {B{\"a}umer}, \citenamefont {Lostaglio}, \citenamefont {Perarnau-Llobet},\
  and\ \citenamefont {Sampaio}}]{baumer2018fluctuating}%
  \BibitemOpen
  \bibfield  {author} {\bibinfo {author} {\bibfnamefont {E.}~\bibnamefont
  {B{\"a}umer}}, \bibinfo {author} {\bibfnamefont {M.}~\bibnamefont
  {Lostaglio}}, \bibinfo {author} {\bibfnamefont {M.}~\bibnamefont
  {Perarnau-Llobet}}, \ and\ \bibinfo {author} {\bibfnamefont {R.}~\bibnamefont
  {Sampaio}},\ }\href@noop {} {\bibfield  {journal} {\bibinfo  {journal}
  {arXiv:1805.10096}\ } (\bibinfo {year} {2018})}\BibitemShut {NoStop}%
\bibitem [{\citenamefont
  {Allahverdyan}(2014)}]{allahverdyan2014nonequilibrium}%
  \BibitemOpen
  \bibfield  {author} {\bibinfo {author} {\bibfnamefont {A.}~\bibnamefont
  {Allahverdyan}},\ }\href@noop {} {\bibfield  {journal} {\bibinfo  {journal}
  {Physical Review E}\ }\textbf {\bibinfo {volume} {90}},\ \bibinfo {pages}
  {032137} (\bibinfo {year} {2014})}\BibitemShut {NoStop}%
\bibitem [{\citenamefont {Solinas}\ and\ \citenamefont
  {Gasparinetti}(2015)}]{solinas2015full}%
  \BibitemOpen
  \bibfield  {author} {\bibinfo {author} {\bibfnamefont {P.}~\bibnamefont
  {Solinas}}\ and\ \bibinfo {author} {\bibfnamefont {S.}~\bibnamefont
  {Gasparinetti}},\ }\href@noop {} {\bibfield  {journal} {\bibinfo  {journal}
  {Physical Review E}\ }\textbf {\bibinfo {volume} {92}},\ \bibinfo {pages}
  {042150} (\bibinfo {year} {2015})}\BibitemShut {NoStop}%
\bibitem [{\citenamefont {Hofer}\ and\ \citenamefont
  {Clerk}(2016)}]{hofer2016negative}%
  \BibitemOpen
  \bibfield  {author} {\bibinfo {author} {\bibfnamefont {P.~P.}\ \bibnamefont
  {Hofer}}\ and\ \bibinfo {author} {\bibfnamefont {A.~A.}\ \bibnamefont
  {Clerk}},\ }\href@noop {} {\bibfield  {journal} {\bibinfo  {journal}
  {Physical Review Letters}\ }\textbf {\bibinfo {volume} {116}},\ \bibinfo
  {pages} {013603} (\bibinfo {year} {2016})}\BibitemShut {NoStop}%
\bibitem [{\citenamefont {Solinas}\ and\ \citenamefont
  {Gasparinetti}(2016)}]{solinas2016probing}%
  \BibitemOpen
  \bibfield  {author} {\bibinfo {author} {\bibfnamefont {P.}~\bibnamefont
  {Solinas}}\ and\ \bibinfo {author} {\bibfnamefont {S.}~\bibnamefont
  {Gasparinetti}},\ }\href@noop {} {\bibfield  {journal} {\bibinfo  {journal}
  {Physical Review A}\ }\textbf {\bibinfo {volume} {94}},\ \bibinfo {pages}
  {052103} (\bibinfo {year} {2016})}\BibitemShut {NoStop}%
\bibitem [{\citenamefont {Solinas}\ \emph {et~al.}(2017)\citenamefont
  {Solinas}, \citenamefont {Miller},\ and\ \citenamefont
  {Anders}}]{solinas2017measurement}%
  \BibitemOpen
  \bibfield  {author} {\bibinfo {author} {\bibfnamefont {P.}~\bibnamefont
  {Solinas}}, \bibinfo {author} {\bibfnamefont {H.}~\bibnamefont {Miller}}, \
  and\ \bibinfo {author} {\bibfnamefont {J.}~\bibnamefont {Anders}},\
  }\href@noop {} {\bibfield  {journal} {\bibinfo  {journal} {Physical Review
  A}\ }\textbf {\bibinfo {volume} {96}},\ \bibinfo {pages} {052115} (\bibinfo
  {year} {2017})}\BibitemShut {NoStop}%
\bibitem [{\citenamefont {Miller}\ and\ \citenamefont
  {Anders}(2017)}]{miller2017time}%
  \BibitemOpen
  \bibfield  {author} {\bibinfo {author} {\bibfnamefont {H.~J.}\ \bibnamefont
  {Miller}}\ and\ \bibinfo {author} {\bibfnamefont {J.}~\bibnamefont
  {Anders}},\ }\href@noop {} {\bibfield  {journal} {\bibinfo  {journal} {New
  Journal of Physics}\ }\textbf {\bibinfo {volume} {19}},\ \bibinfo {pages}
  {062001} (\bibinfo {year} {2017})}\BibitemShut {NoStop}%
\bibitem [{\citenamefont {Miller}\ and\ \citenamefont
  {Anders}(2018)}]{miller2018leggett}%
  \BibitemOpen
  \bibfield  {author} {\bibinfo {author} {\bibfnamefont {H.~J.}\ \bibnamefont
  {Miller}}\ and\ \bibinfo {author} {\bibfnamefont {J.}~\bibnamefont
  {Anders}},\ }\href@noop {} {\bibfield  {journal} {\bibinfo  {journal}
  {Entropy}\ }\textbf {\bibinfo {volume} {20}},\ \bibinfo {pages} {200}
  (\bibinfo {year} {2018})}\BibitemShut {NoStop}%
\bibitem [{\citenamefont {{\AA}berg}(2018)}]{aberg2016fully}%
  \BibitemOpen
  \bibfield  {author} {\bibinfo {author} {\bibfnamefont {J.}~\bibnamefont
  {{\AA}berg}},\ }\href@noop {} {\bibfield  {journal} {\bibinfo  {journal}
  {Physical Review X}\ }\textbf {\bibinfo {volume} {8}},\ \bibinfo {pages}
  {011019} (\bibinfo {year} {2018})}\BibitemShut {NoStop}%
\bibitem [{\citenamefont {Sampaio}\ \emph {et~al.}(2018)\citenamefont
  {Sampaio}, \citenamefont {Suomela}, \citenamefont {Ala-Nissila},
  \citenamefont {Anders},\ and\ \citenamefont {Philbin}}]{sampaio2018quantum}%
  \BibitemOpen
  \bibfield  {author} {\bibinfo {author} {\bibfnamefont {R.}~\bibnamefont
  {Sampaio}}, \bibinfo {author} {\bibfnamefont {S.}~\bibnamefont {Suomela}},
  \bibinfo {author} {\bibfnamefont {T.}~\bibnamefont {Ala-Nissila}}, \bibinfo
  {author} {\bibfnamefont {J.}~\bibnamefont {Anders}}, \ and\ \bibinfo {author}
  {\bibfnamefont {T.}~\bibnamefont {Philbin}},\ }\href@noop {} {\bibfield
  {journal} {\bibinfo  {journal} {Physical Review A}\ }\textbf {\bibinfo
  {volume} {97}},\ \bibinfo {pages} {012131} (\bibinfo {year}
  {2018})}\BibitemShut {NoStop}%
\bibitem [{\citenamefont {Francica}\ \emph {et~al.}(2017)\citenamefont
  {Francica}, \citenamefont {Goold},\ and\ \citenamefont
  {Plastina}}]{francica2017role}%
  \BibitemOpen
  \bibfield  {author} {\bibinfo {author} {\bibfnamefont {G.}~\bibnamefont
  {Francica}}, \bibinfo {author} {\bibfnamefont {J.}~\bibnamefont {Goold}}, \
  and\ \bibinfo {author} {\bibfnamefont {F.}~\bibnamefont {Plastina}},\
  }\href@noop {} {\bibfield  {journal} {\bibinfo  {journal} {arXiv:1707.06950}\
  } (\bibinfo {year} {2017})}\BibitemShut {NoStop}%
\bibitem [{\citenamefont {Xu}\ \emph {et~al.}(2018)\citenamefont {Xu},
  \citenamefont {Zou}, \citenamefont {Guo},\ and\ \citenamefont
  {Kong}}]{xu2018effects}%
  \BibitemOpen
  \bibfield  {author} {\bibinfo {author} {\bibfnamefont {B.-M.}\ \bibnamefont
  {Xu}}, \bibinfo {author} {\bibfnamefont {J.}~\bibnamefont {Zou}}, \bibinfo
  {author} {\bibfnamefont {L.-S.}\ \bibnamefont {Guo}}, \ and\ \bibinfo
  {author} {\bibfnamefont {X.-M.}\ \bibnamefont {Kong}},\ }\href@noop {}
  {\bibfield  {journal} {\bibinfo  {journal} {Physical Review A}\ }\textbf
  {\bibinfo {volume} {97}},\ \bibinfo {pages} {052122} (\bibinfo {year}
  {2018})}\BibitemShut {NoStop}%
\bibitem [{\citenamefont {Xu}\ \emph {et~al.}(2019)\citenamefont {Xu},
  \citenamefont {Tu}, \citenamefont {Zou},\ and\ \citenamefont
  {Wang}}]{xu2019duality}%
  \BibitemOpen
  \bibfield  {author} {\bibinfo {author} {\bibfnamefont {B.-M.}\ \bibnamefont
  {Xu}}, \bibinfo {author} {\bibfnamefont {Z.-C.}\ \bibnamefont {Tu}}, \bibinfo
  {author} {\bibfnamefont {J.}~\bibnamefont {Zou}}, \ and\ \bibinfo {author}
  {\bibfnamefont {J.}~\bibnamefont {Wang}},\ }\href@noop {} {\bibfield
  {journal} {\bibinfo  {journal} {arXiv:1903.01064}\ } (\bibinfo {year}
  {2019})}\BibitemShut {NoStop}%
\bibitem [{\citenamefont {Wu}\ \emph {et~al.}(2019)\citenamefont {Wu},
  \citenamefont {Yuan}, \citenamefont {Xiang}, \citenamefont {Li},
  \citenamefont {Guo},\ and\ \citenamefont
  {Perarnau-Llobet}}]{wu2019experimentally}%
  \BibitemOpen
  \bibfield  {author} {\bibinfo {author} {\bibfnamefont {K.-D.}\ \bibnamefont
  {Wu}}, \bibinfo {author} {\bibfnamefont {Y.}~\bibnamefont {Yuan}}, \bibinfo
  {author} {\bibfnamefont {G.-Y.}\ \bibnamefont {Xiang}}, \bibinfo {author}
  {\bibfnamefont {C.-F.}\ \bibnamefont {Li}}, \bibinfo {author} {\bibfnamefont
  {G.-C.}\ \bibnamefont {Guo}}, \ and\ \bibinfo {author} {\bibfnamefont
  {M.}~\bibnamefont {Perarnau-Llobet}},\ }\href@noop {} {\bibfield  {journal}
  {\bibinfo  {journal} {Science advances}\ }\textbf {\bibinfo {volume} {5}},\
  \bibinfo {pages} {eaav4944} (\bibinfo {year} {2019})}\BibitemShut {NoStop}%
\bibitem [{\citenamefont {Allahverdyan}\ and\ \citenamefont
  {Nieuwenhuizen}(2005)}]{allahverdyan2005fluctuations}%
  \BibitemOpen
  \bibfield  {author} {\bibinfo {author} {\bibfnamefont {A.}~\bibnamefont
  {Allahverdyan}}\ and\ \bibinfo {author} {\bibfnamefont {T.~M.}\ \bibnamefont
  {Nieuwenhuizen}},\ }\href@noop {} {\bibfield  {journal} {\bibinfo  {journal}
  {Physical Review E}\ }\textbf {\bibinfo {volume} {71}},\ \bibinfo {pages}
  {066102} (\bibinfo {year} {2005})}\BibitemShut {NoStop}%
\bibitem [{\citenamefont {Margenau}\ and\ \citenamefont
  {Hill}(1961)}]{margenau1961correlation}%
  \BibitemOpen
  \bibfield  {author} {\bibinfo {author} {\bibfnamefont {H.}~\bibnamefont
  {Margenau}}\ and\ \bibinfo {author} {\bibfnamefont {R.~N.}\ \bibnamefont
  {Hill}},\ }\href@noop {} {\bibfield  {journal} {\bibinfo  {journal} {Progress
  of Theoretical Physics}\ }\textbf {\bibinfo {volume} {26}},\ \bibinfo {pages}
  {722} (\bibinfo {year} {1961})}\BibitemShut {NoStop}%
\bibitem [{\citenamefont {Nazarov}\ and\ \citenamefont
  {Kindermann}(2003)}]{nazarov2003full}%
  \BibitemOpen
  \bibfield  {author} {\bibinfo {author} {\bibfnamefont {Y.~V.}\ \bibnamefont
  {Nazarov}}\ and\ \bibinfo {author} {\bibfnamefont {M.}~\bibnamefont
  {Kindermann}},\ }\href@noop {} {\bibfield  {journal} {\bibinfo  {journal}
  {The European Physical Journal B-Condensed Matter and Complex Systems}\
  }\textbf {\bibinfo {volume} {35}},\ \bibinfo {pages} {413} (\bibinfo {year}
  {2003})}\BibitemShut {NoStop}%
\bibitem [{\citenamefont {Johansen}(2007)}]{johansen2007quantum}%
  \BibitemOpen
  \bibfield  {author} {\bibinfo {author} {\bibfnamefont {L.~M.}\ \bibnamefont
  {Johansen}},\ }\href@noop {} {\bibfield  {journal} {\bibinfo  {journal}
  {Physical Review A}\ }\textbf {\bibinfo {volume} {76}},\ \bibinfo {pages}
  {012119} (\bibinfo {year} {2007})}\BibitemShut {NoStop}%
\bibitem [{\citenamefont {Lundeen}\ and\ \citenamefont
  {Bamber}(2012)}]{lundeen2012procedure}%
  \BibitemOpen
  \bibfield  {author} {\bibinfo {author} {\bibfnamefont {J.~S.}\ \bibnamefont
  {Lundeen}}\ and\ \bibinfo {author} {\bibfnamefont {C.}~\bibnamefont
  {Bamber}},\ }\href@noop {} {\bibfield  {journal} {\bibinfo  {journal}
  {Physical Review Letters}\ }\textbf {\bibinfo {volume} {108}},\ \bibinfo
  {pages} {070402} (\bibinfo {year} {2012})}\BibitemShut {NoStop}%
\bibitem [{\citenamefont {Wigner}(1932)}]{wigner1997quantum}%
  \BibitemOpen
  \bibfield  {author} {\bibinfo {author} {\bibfnamefont {E.}~\bibnamefont
  {Wigner}},\ }\href {\doibase 10.1103/PhysRev.40.749} {\bibfield  {journal}
  {\bibinfo  {journal} {Phys. Rev.}\ }\textbf {\bibinfo {volume} {40}},\
  \bibinfo {pages} {749} (\bibinfo {year} {1932})}\BibitemShut {NoStop}%
\bibitem [{\citenamefont {Hillery}\ \emph {et~al.}(1984)\citenamefont
  {Hillery}, \citenamefont {O'Connell}, \citenamefont {Scully},\ and\
  \citenamefont {Wigner}}]{hillery1984distribution}%
  \BibitemOpen
  \bibfield  {author} {\bibinfo {author} {\bibfnamefont {M.}~\bibnamefont
  {Hillery}}, \bibinfo {author} {\bibfnamefont {R.~F.}\ \bibnamefont
  {O'Connell}}, \bibinfo {author} {\bibfnamefont {M.~O.}\ \bibnamefont
  {Scully}}, \ and\ \bibinfo {author} {\bibfnamefont {E.~P.}\ \bibnamefont
  {Wigner}},\ }\href@noop {} {\bibfield  {journal} {\bibinfo  {journal}
  {Physics reports}\ }\textbf {\bibinfo {volume} {106}},\ \bibinfo {pages}
  {121} (\bibinfo {year} {1984})}\BibitemShut {NoStop}%
\bibitem [{\citenamefont {Polkovnikov}(2010)}]{polkovnikov2010phase}%
  \BibitemOpen
  \bibfield  {author} {\bibinfo {author} {\bibfnamefont {A.}~\bibnamefont
  {Polkovnikov}},\ }\href@noop {} {\bibfield  {journal} {\bibinfo  {journal}
  {Annals of Physics}\ }\textbf {\bibinfo {volume} {325}},\ \bibinfo {pages}
  {1790} (\bibinfo {year} {2010})}\BibitemShut {NoStop}%
\bibitem [{\citenamefont {Kac}(1949)}]{kac1949distributions}%
  \BibitemOpen
  \bibfield  {author} {\bibinfo {author} {\bibfnamefont {M.}~\bibnamefont
  {Kac}},\ }\href@noop {} {\bibfield  {journal} {\bibinfo  {journal}
  {Transactions of the American Mathematical Society}\ }\textbf {\bibinfo
  {volume} {65}},\ \bibinfo {pages} {1} (\bibinfo {year} {1949})}\BibitemShut
  {NoStop}%
\bibitem [{\citenamefont {Liu}(2012)}]{liu2012derivation}%
  \BibitemOpen
  \bibfield  {author} {\bibinfo {author} {\bibfnamefont {F.}~\bibnamefont
  {Liu}},\ }\href@noop {} {\bibfield  {journal} {\bibinfo  {journal} {Physical
  Review E}\ }\textbf {\bibinfo {volume} {86}},\ \bibinfo {pages} {010103}
  (\bibinfo {year} {2012})}\BibitemShut {NoStop}%
\bibitem [{\citenamefont {Talkner}\ \emph {et~al.}(2008)\citenamefont
  {Talkner}, \citenamefont {Burada},\ and\ \citenamefont
  {H{\"a}nggi}}]{talkner2008statistics}%
  \BibitemOpen
  \bibfield  {author} {\bibinfo {author} {\bibfnamefont {P.}~\bibnamefont
  {Talkner}}, \bibinfo {author} {\bibfnamefont {P.~S.}\ \bibnamefont {Burada}},
  \ and\ \bibinfo {author} {\bibfnamefont {P.}~\bibnamefont {H{\"a}nggi}},\
  }\href@noop {} {\bibfield  {journal} {\bibinfo  {journal} {Physical Review
  E}\ }\textbf {\bibinfo {volume} {78}},\ \bibinfo {pages} {011115} (\bibinfo
  {year} {2008})}\BibitemShut {NoStop}%
\bibitem [{\citenamefont {Deffner}\ and\ \citenamefont
  {Lutz}(2008)}]{deffner2008nonequilibrium}%
  \BibitemOpen
  \bibfield  {author} {\bibinfo {author} {\bibfnamefont {S.}~\bibnamefont
  {Deffner}}\ and\ \bibinfo {author} {\bibfnamefont {E.}~\bibnamefont {Lutz}},\
  }\href@noop {} {\bibfield  {journal} {\bibinfo  {journal} {Physical Review
  E}\ }\textbf {\bibinfo {volume} {77}},\ \bibinfo {pages} {021128} (\bibinfo
  {year} {2008})}\BibitemShut {NoStop}%
\bibitem [{\citenamefont {Pan}\ \emph {et~al.}(2018)\citenamefont {Pan},
  \citenamefont {Hoang}, \citenamefont {Fei}, \citenamefont {Qiu},
  \citenamefont {Ahn}, \citenamefont {Li},\ and\ \citenamefont
  {Quan}}]{pan2018quantifying}%
  \BibitemOpen
  \bibfield  {author} {\bibinfo {author} {\bibfnamefont {R.}~\bibnamefont
  {Pan}}, \bibinfo {author} {\bibfnamefont {T.~M.}\ \bibnamefont {Hoang}},
  \bibinfo {author} {\bibfnamefont {Z.}~\bibnamefont {Fei}}, \bibinfo {author}
  {\bibfnamefont {T.}~\bibnamefont {Qiu}}, \bibinfo {author} {\bibfnamefont
  {J.}~\bibnamefont {Ahn}}, \bibinfo {author} {\bibfnamefont {T.}~\bibnamefont
  {Li}}, \ and\ \bibinfo {author} {\bibfnamefont {H.~T.}\ \bibnamefont
  {Quan}},\ }\href@noop {} {\bibfield  {journal} {\bibinfo  {journal} {Physical
  Review E}\ }\textbf {\bibinfo {volume} {98}},\ \bibinfo {pages} {052105}
  (\bibinfo {year} {2018})}\BibitemShut {NoStop}%
\bibitem [{\citenamefont {Barnett}\ and\ \citenamefont
  {Radmore}(2002)}]{barnett2002methods}%
  \BibitemOpen
  \bibfield  {author} {\bibinfo {author} {\bibfnamefont {S.}~\bibnamefont
  {Barnett}}\ and\ \bibinfo {author} {\bibfnamefont {P.~M.}\ \bibnamefont
  {Radmore}},\ }\href@noop {} {\emph {\bibinfo {title} {Methods in theoretical
  quantum optics}}},\ Vol.~\bibinfo {volume} {15}\ (\bibinfo  {publisher}
  {Oxford University Press},\ \bibinfo {year} {2002})\BibitemShut {NoStop}%
\end{thebibliography}%

%

\end{document}